# Chapter 2

# Machine Layout and Performance

D. Angal-Kalinin[1], R. Appleby[2], G. Arduini[3]*, D. Banfi[4], J. Barranco[4], N. Biancacci[3], D. Brett[2], R. Bruce[3], O. Brüning[3], X. Buffat[3], A. Burov[5], Y. Cai[6], R. Calaga[3], A. Chancé[7], M. Crouch[2], B. Dalena[7], H. Day[3], R. de Maria[3], J. Esteban Muller[3], S. Fartoukh[3], M. Fitterer[3], O. Frasciello[8], M. Giovannozzi[3], W. Herr[3], W. Höfle[3], B. Holzer[3], G. Iadarola[3], J.M. Jowett[3], M. Korostelev[9], K. Li[3], E. McIntosh[3], E. Métral[3], A. Mostacci[10], N. Mounet[3], B. Muratori[1], Y. Nosochkov[6], K. Ohmi[11], Y. Papaphilippou[3], S. Paret[12], J. Payet[7], T. Pieloni[3], J. Qiang[12], T. Rijoff[3], L. Rossi[3], G. Rumolo[3], B. Salvant[3], M. Schaumann[3], E. Shaposhnikova[3], D. Shatilov[13], C. Tambasco[3], R. Tomás[3], A. Valishev[5], M.-H. Wang[6], R. Wanzenberg[14], S. White[3], A. Wolski[9], O. Zagorodnova[14], C. Zannini[3], F. Zimmermann[3] and M. Zobov[8]

[1]ASTeC, STFC, Daresbury Laboratory, Warrington, UK
[2]UMAN, The University of Manchester and the Cockcroft Institute, Warrington, UK
[3]CERN, Accelerator & Technology Sector, Geneva, Switzerland
[4]EPFL, Lausanne, Switzerland
[5]FNAL, Fermi National Accelerator Laboratory, Batavia, USA
[6]SLAC, National Accelerator Laboratory, Menlo Park, USA
[7]CEA/SACLAY, DSM/Irfu/SACM, Gif-sur-Yvette, France
[8]INFN-LNF, Rome, Italy
[9]University of Liverpool, Liverpool, UK
[10]University of Rome "La Sapienza", Rome, Italy
[11]KEK, Tsukuba, Japan
[12]LBNL, Lawrence Berkeley National Laboratory, Berkeley, USA
[13]BINP, Novosibirsk, Russia
[14]DESY, Deutsches Elektronen-Synchrotron, Hamburg, Germany

## 2    Machine layout and performance

### 2.1   Performance goals (nominal scheme)

The goal of the High Luminosity upgrade of the LHC is to deliver an integrated luminosity of at least 250 fb$^{-1}$ per year in each of the two high-luminosity general-purpose detectors, ATLAS and CMS, located at the interaction points (IP) one and five, respectively. The other two experiments, ALICE and LHCb with detectors located at IP2 and IP8 respectively, are expecting to collect integrated luminosities of 100 pb$^{-1}$ per year (of proton–proton data) and 5 fb$^{-1}$ to 10 fb$^{-1}$ per year, respectively [1]. No operation for forward physics experiments is expected after the upgrade.

The ATLAS and CMS detectors will be upgraded to handle an average number of pile-up events per bunch crossing of at least 140, corresponding to an instantaneous luminosity of approximately $5 \times 10^{34}$ cm$^{-2}$ s$^{-1}$ for operation with 25 ns beams at 7 TeV, for a visible cross-section $\sigma_{vis}$ = 85 mb. The detectors are also expected to handle a line density of pile-up events of 1.3 events per mm per bunch crossing. ALICE and LHCb will be upgraded to operate at instantaneous luminosities of up to $2 \times 10^{31}$ cm$^{-2}$ s$^{-1}$ and $2 \times 10^{33}$ cm$^{-2}$ s$^{-1}$, respectively.

The HL-LHC upgrade project aims to achieve a 'virtual' peak luminosity that is considerably higher than the maximum imposed by the acceptable event pile-up rate, and to control the instantaneous luminosity during the physics fill ('luminosity levelling') so that the luminosity production can be sustained over longer periods to maximize the integrated luminosity.

---

* Corresponding author: Gianluigi.Arduini@cern.ch

A simplified but realistic model of the luminosity evolution has been developed [2] taking into account the beam population $N_{\text{beam}}$ reduction due to the collisions (the so called 'burn-off') in $n_{\text{IP}}$ collision points with instantaneous luminosity $L_{\text{inst}}$,

$$\frac{dN_{\text{beam}}}{dt} = -n_{\text{IP}}\sigma_{\text{tot}}L_{\text{inst}}, \quad (2\text{-}1)$$

where $\sigma_{\text{tot}}$ is the total hadron cross-section (here assumed to be 100 mb). No other sources of intensity reduction or emittance blow-up are considered in this model. Figure 2-1 shows the expected yearly integrated luminosity as a function of the 'virtual' peak luminosity for three different values of the luminosity at which levelling is performed (see Section 1.2.3). In this figure the corresponding optimum fill length $T_{\text{fill}}$ (i.e. the length of time for each fill that will maximize the average luminosity production rate) is also shown. In order to estimate the annual integrated luminosity, we assume a minimum turnaround time $T_{\text{turnaround}}$ of 3 hours (see Chapter 16), a scheduled physics time $T_{\text{physics}}$ for luminosity production of 160 days per year, with $N_{\text{fills}}$ successful physics fills of duration $T_{\text{fill}}$, and a performance efficiency of 50% (this was 53.5% in 2012) where [3]:

$$\eta = N_{\text{fills}} \frac{T_{\text{turnaround}} + T_{\text{fill}}}{T_{\text{physics}}} \times 100\% \quad (2\text{-}2)$$

In order to reach the goal of integrating 250 fb$^{-1}$/year levelling must be performed at luminosities larger than $5 \times 10^{34}$ cm$^{-2}$ s$^{-1}$ and peak virtual luminosities of more than $20 \times 10^{34}$ cm$^{-2}$ s$^{-1}$. Furthermore, the performance efficiency must be at least 50% and the typical fill length must be comparable with the estimated optimum fill length (for comparison the average fill length during the 2012 run was 6.1 hours). In this respect, levelling to higher luminosities will be beneficial because it would make it easier to reach and even exceed the integrated luminosity goal, with fill lengths comparable to the fill lengths of the 2012 run.

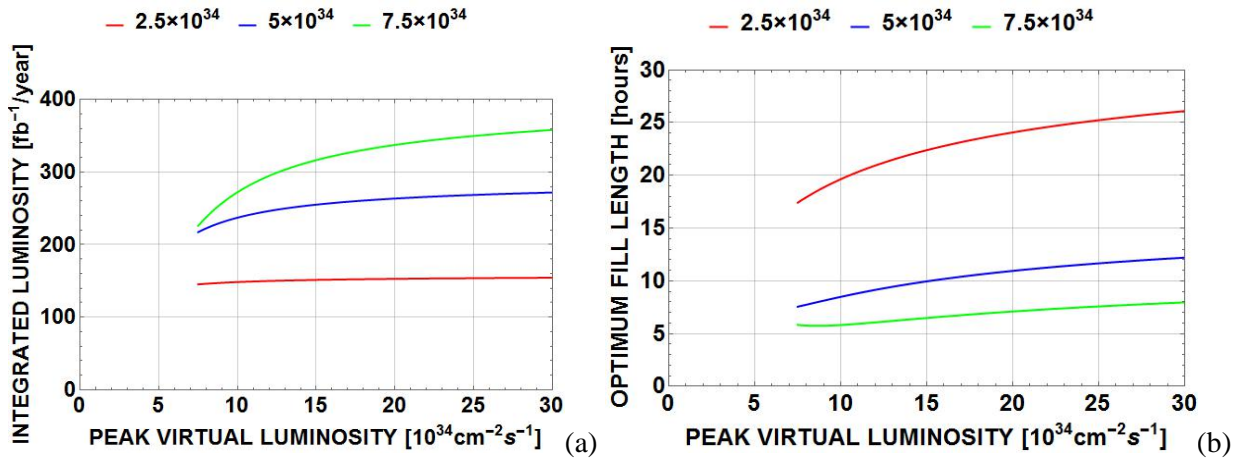

Figure 2-1: (a) Expected annual integrated luminosity; (b) optimum fill length as a function of the 'virtual' peak luminosity for three different values of the luminosity at which levelling is performed. A circulating current of 1.1 A (corresponding to $N_{\text{beam}} = 6.1 \times 10^{14}$ p), a minimum turnaround time of 3 hours and a performance efficiency $\eta$ of 50% have been assumed. Only burn-off for a total hadron cross-section of 100 mb has been considered for the estimate of the beam population and virtual luminosity evolution. Two high-luminosity interaction points have been assumed.

2.1.1 Parameter space and basic parameter choices

The instantaneous luminosity $L$ is given by

$$L = \frac{n_b N^2 f_{\text{rev}} \gamma}{4\pi\beta^* \varepsilon_n} R(\beta^*, \sigma_z, d_{\text{bb}}) \quad (2\text{-}3)$$

The r.m.s. normalized emittance $\varepsilon_n$ in collision is assumed here to be equal for the two beams and for the horizontal and vertical planes. The Twiss beta function $\beta^*$ in collision at the IP determines, together with the normalized emittance, the r.m.s. beam size $\sigma^* = \sqrt{\varepsilon_n \beta^*/\gamma}$ at the IP (assuming that the contribution to the beam size due to the dispersion and the momentum spread of the beam can be neglected). Here and below it is assumed that the relativistic factor $\beta = 1$.

A crossing angle is needed to separate bunches immediately upstream and downstream of the collision point. This leads to a reduced geometric overlap between the colliding beams, and hence to a reduction in luminosity. The crossing angle needs to be increased when reducing the $\beta^*$ in order to maintain a sufficiently large normalized beam–beam separation $d_{bb}$. The luminosity is also reduced by the 'hourglass effect' that arises from the increase of the beta function upstream and downstream of the interaction point along the bunch longitudinal distribution. The hourglass effect is enhanced by a reduction in $\beta^*$ and by an increase in bunch length. The luminosity reduction factor $R$ in Eq. (2-3) takes both the crossing angle and the hourglass effect into account.

Equation (2-3) shows the parameters that can be varied to maximize the instantaneous luminosity. The considerations that constrain their values are briefly discussed below [4, 5]:

- The maximum number of bunches nb is limited by the minimum time interval between bunch crossings at the IP that can be handled by the detectors: this is limited to 25 ns. The maximum number of bunches that can be injected in the LHC is also limited by the following.

    o   The maximum number of bunches that can be transferred safely from the SPS to the LHC due to the maximum energy that can be deposited on the injection protection absorber (TDI) in case the LHC injection kicker is not firing. The present limitation for the TDI is considered to be a maximum of 288 bunches per SPS extraction for the ultimate bunch population [6].

    o   The rise-time of the injection kickers in the SPS and LHC, extraction kickers in the PS and SPS, and abort gap kicker in the LHC.

    o   The need for injecting one train consisting of a few bunches (typically 12 nominal bunches for 25 ns spacing) before injecting one nominal train for machine protection considerations [7]. For the same reason the last train must have the maximum number of bunches.

    o   The constraints imposed by the experiments: the need for non-colliding bunches for background evaluation, and a sufficient number of collisions for lower luminosity experiments [1].

- The maximum bunch population N is limited in the LHC by the onset of the single bunch transverse mode coupling instability (TMCI), expected to occur at $3.5 \times 10^{11}$ p/bunch [8].

- The total current of the beam circulating in the LHC, $I_{\text{beam}} = e n_b N f_{\text{rev}}$ (where e is the proton charge), is expected to be limited to 1.1 A by the cryogenic power available to cool the beam screen. This assumes that a secondary electron yield (SEY) as low as 1.3 can be reached in the beam screen, to limit the heat load due to the electron cloud in the arcs, and additional cryogenic plants are installed in Points 1, 4 and 5 [4, 9].

- The beam brightness $B \equiv N/\varepsilon_n$ is limited by the following considerations [4].

    o   The total head-on beam–beam tune shift $\Delta Q_{\text{bbho}} \propto N/\varepsilon_n$ is expected to be limited to 0.02–0.03 based on experience gained (from operations and dedicated experiments) during LHC Run 1. Its value is reduced in a similar fashion to the luminosity in the presence of a crossing angle [10].

    o   Intra-beam scattering induces transverse and longitudinal emittance blow-up, particularly at injection (low energy) but also in the acceleration, squeeze, and collision phases. The evolution of the beam emittances can be described by the equations,
    $$\frac{1}{\tau_H} = \frac{1}{\varepsilon_{nH}} \frac{d\varepsilon_{nH}}{dt} \text{ and } \frac{1}{\tau_L} = \frac{1}{\varepsilon_L} \frac{d\varepsilon_L}{dt} \text{ with } \frac{1}{\tau_d} \propto \frac{N}{\gamma \varepsilon_{nH} \varepsilon_{nV} \varepsilon_L} \text{ and } d = \text{H, L,} \qquad (2\text{-}4)$$

where $\varepsilon_{nH,V}$ are the r.m.s. normalized horizontal and vertical emittances. Here we assume that vertical dispersion and coupling are negligible so that the vertical emittance blow-up can be neglected.

The minimum $\beta^*$ is limited by [5]:

- The aperture at the triplet, taking into account that the maximum $\beta$ function $\beta_{max}$ at the triplet increases in inverse proportion to $\beta^*$, and that the crossing angle $\theta_c$ required to maintain a sufficiently large normalized beam–beam separation $d_{bb}$ to minimize the long-range beam–beam tune spread $\Delta Q_{bbLR}$ is $\theta_c = d_{bb}\sqrt{\varepsilon_n/\gamma\beta^*}$;
- The maximum $\beta$ function at the triplet that can be matched to the regular optics of the arcs within the distance available in the matching section between the triplets and the arcs;
- The strengths of the arc sextupoles available to correct the chromaticity generated by the triplets (proportional to $\beta_{max}$) and, in general, the nonlinear chromaticities and off-momentum beta beating.

For a round optics (i.e. with equal $\beta^*$ in the horizontal and vertical planes) in the presence of a crossing angle and at constant normalized long-range beam–beam separation $d_{bb}$, the increase in luminosity saturates for values of $\beta^* < \sigma_z$, as shown in Figure 2-2, because of the corresponding reduction of the luminosity reduction factor $R$. The effect of the geometric reduction due to the crossing angle can be counteracted by means of crab cavities operated at the LHC main RF frequency [11] as shown in Figure 1-6. The comparison of the two plots of Figure 2-2 also shows that the effect of crab cavities in enhancing the peak virtual luminosity becomes negligible, for $\beta^*$ greater than 30–40 cm.

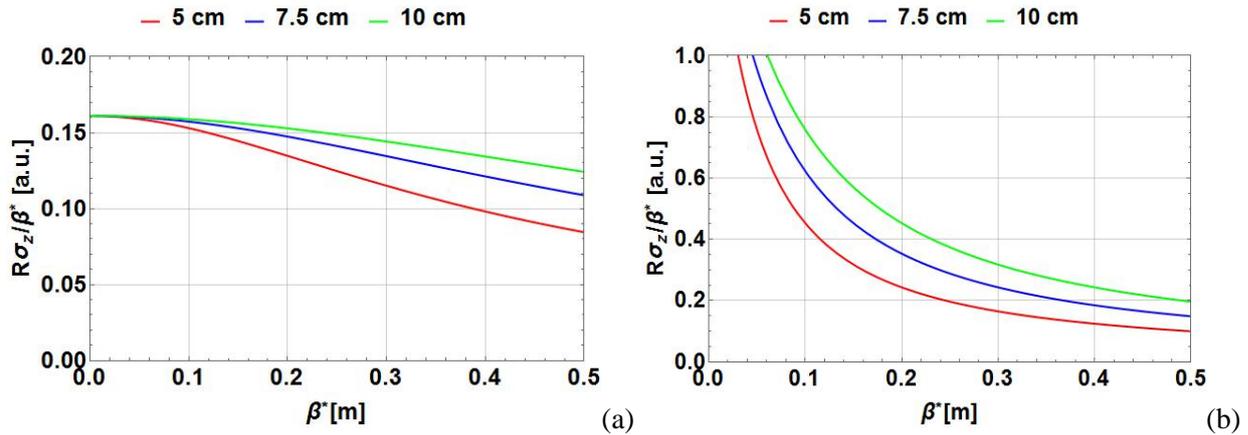

Figure 2-2: Parameter $R\sigma_z/\beta^*$ vs. $\beta^*$ for different bunch lengths for a round optics and constant normalized long-range beam–beam separation $d_{bb}$ (a) without crab cavities and (b) with crab cavities. The small effect of RF curvature in the crab cavities is not included.

Even after their planned upgrades, the injectors will also constrain the parameters of the beam that can be expected in the LHC in collision. First, there is a maximum current $I_{RF} \approx 2N_{SPS}e/T_{bb}$ (where $N_{SPS}$ is the bunch population in the SPS and $T_{bb}$ is the bunch spacing) that can be accelerated per SPS cycle due to RF power limitations in the power amplifiers, power couplers and feeder lines of the main 200 MHz RF system. This maximum current is 2.6 A corresponding to $N_{SPS} = 2.0 \times 10^{11}$ particles at SPS extraction [12, 13]. The total number of bunches is also limited to 288 due to the thermal load on the power lines. Second, the brightness of the LHC beam in the injectors is expected to be limited by space charge effects at injection in the PSB, PS, and SPS. From present experience it is expected that the maximum brightness of the LHC beams after the full injector upgrade will be $B_{SPS} \sim 1.5 \times 10^{11}$ p/μm [13].

Table 2-1 shows the beam parameters at collision, selected on the basis of the above considerations [3, 14]. The parameters in the table are consistent with the above constraints with the exception of the requested bunch population that can be delivered within the longitudinal acceptance of the LHC at injection. To avoid longitudinal instabilities in the SPS, a controlled longitudinal emittance blow-up needs to be applied for high bunch population, which would lead to bunches that are longer than acceptable for clean capture in the LHC with the main 400 MHz RF system (even with the maximum 200 MHz RF voltage available in the SPS at extraction). The identification of the elements contributing to the longitudinal impedance in the SPS and the reduction of their impedance might allow for this limit to be increased. In that case, the requirements in terms of bunch population for the HL-LHC could be met.

Table 2-1: HL-LHC nominal parameters for 25 ns operation [14] for two production modes of the LHC beam in the injectors described in Ref. [3].

| Parameter | Nominal LHC (design report) | HL-LHC (standard) | HL-LHC (BCMS) |
|---|---|---|---|
| Beam energy in collision [TeV] | 7 | 7 | 7 |
| Particles per bunch, $N$ [$10^{11}$] | 1.15 | 2.2 | 2.2 |
| Number of bunches per beam | 2808 | 2748 | 2604 |
| Number of collisions in IP1 and IP5[*] | 2808 | 2736 | 2592 |
| $N_{tot}$ [$10^{14}$] | 3.2 | 6.0 | 5.7 |
| Beam current [A] | 0.58 | 1.09 | 1.03 |
| Crossing angle in IP1 and IP5 [μrad] | 285 | 590 | 590 |
| Normalized long-range beam–beam separation [$\sigma$] | 9.4 | 12.5 | 12.5 |
| Minimum $\beta^*$ [m] | 0.55 | 0.15 | 0.15 |
| $\varepsilon_n$ [μm] | 3.75 | 2.50 | 2.50 |
| $\varepsilon_L$ [eVs] | 2.50 | 2.50 | 2.50 |
| r.m.s. energy spread [0.0001] | 1.13 | 1.13 | 1.13 |
| r.m.s. bunch length [cm] | 7.55 | 7.55 | 7.55 |
| IBS horizontal [h] | 105 | 18.5 | 18.5 |
| IBS longitudinal [h] | 63 | 20.4 | 20.4 |
| Piwinski parameter | 0.65 | 3.14 | 3.14 |
| Total loss factor $R_0$ without crab cavity | 0.836 | 0.305 | 0.305 |
| Total loss factor $R_1$ with crab cavity | (0.981) | 0.829 | 0.829 |
| Beam–beam/IP without crab cavity | 0.0031 | 0.0033 | 0.0033 |
| Beam–beam/IP with crab cavity | 0.0038 | 0.011 | 0.011 |
| Peak luminosity without crab cavity [$10^{34}$ cm$^{-2}$ s$^{-1}$] | 1.00 | 7.18 | 6.80 |
| Virtual luminosity with crab cavity $L_{peak} \times R_1/R_0$ [$10^{34}$ cm$^{-2}$ s$^{-1}$] | (1.18) | 19.54 | 18.52 |
| Events/crossing without levelling and without crab cavity | 27 | 198 | 198 |
| Levelled luminosity [$10^{34}$ cm$^{-2}$ s$^{-1}$] | - | 5.00[†] | 5.00[†] |
| Events/crossing (with levelling and crab cavities for HL-LHC)[‡] | 27 | 138 | 146 |
| Maximum line density of pile-up events during fill [event/mm] | 0.21 | 1.25 | 1.31 |
| Levelling time [h] (assuming no emittance growth)[‡] | - | 8.3 | 7.6 |
| Number of collisions in IP2/IP8 | 2808 | 2452/2524[**] | 2288/2396[**] |
| $N$ at LHC injection [$10^{11}$][††] | 1.20 | 2.30 | 2.30 |
| Maximum number of bunches per injection | 288 | 288 | 288 |
| $N_{tot}$/injection [$10^{13}$] | 3.46 | 6.62 | 6.62 |
| $\varepsilon_n$ at SPS extraction [μm][‡‡] | 3.40 | 2.00 | <2.00[***] |

*Assuming one less batch from the PS for machine protection (pilot injection, Transfer line steering with 12 nominal bunches) and non-colliding bunches for experiments (background studies, etc.). Note that due to RF beam loading the abort gap length must not exceed the 3 μs design value.
†For the design of the HL-LHC systems (collimators, triplet magnets, etc.), a margin of 50% on the stated peak luminosity (corresponding to the ultimate levelled luminosity) has been agreed.
‡The total number of events/crossing is calculated with an inelastic cross-section of 85 mb (also for nominal), while 100 mb is still assumed for calculating the proton burn off and the resulting levelling time.
**The lower number of collisions in IR2/8 compared to the general-purpose detectors is a result of the agreed filling scheme, aiming as much as possible at a democratic sharing of collisions between the experiments.
††An intensity loss of 5% distributed along the cycle is assumed from SPS extraction to collisions in the LHC.
‡‡A transverse emittance blow-up of 10–15% on the average H/V emittance in addition to that expected from intra-beam scattering (IBS) is assumed (to reach 2.5 μm of emittance in collision for 25 ns operation).
***For the BCMS scheme emittances down to 1.7 μm have already been achieved at LHC injection, which might be used to mitigate excessive emittance blow-up in the LHC during injection and ramp.

## 2.2 Proposed systems upgrades and improvements

The high luminosity configuration requires upgrades of numerous systems. In some cases, existing systems would not be able to face the increasingly harsh conditions that the highest luminosity performance will generate. Accelerated wear and radiation damage are serious concerns. Many changes will be necessary just in order to allow the machine to continue to run in a regime of nominal or ultimate luminosity. For certain systems, replacements could be made with equipment achieving better performance, rather than with spares of the same specification. This performance 'improvement' goes well beyond the basic consolidation that is already planned for the LHC.

For other systems, replacement, although triggered by technical reasons, is the chance to carry out a complete change of layout or performance and may be considered to be a real upgrade. The most striking example is the replacement of the inner triplet magnets with new magnets of different technology based on a $Nb_3Sn$ superconductor. This will constitute the backbone of the upgrade. Another case is the replacement of a good part of the present collimation system with an improved design with lower impedance jaws.

In other cases, new equipment not included in the present LHC layout will be installed in order to increase performance, either in terms of peak luminosity or availability. The most important example is the superconducting RF crab cavities, which are of a compact design as required for the HL-LHC, comprising a completely new development and a first for a proton collider. A further example is the installation of a collimation system in the continuous cryostat in the dispersion suppressors.

In this section, we compile a list of the systems that will require an upgrade or at least a serious improvement in performance, to face the ambitious challenge of the High Luminosity LHC.

### 2.2.1 Insertion region magnets

It is expected that the LHC will reach an integrated luminosity of approximately 300 fb$^{-1}$ by 2022, resulting in doses of up to 30 MGy to some components in the high luminosity interaction regions. The inner triplet quadrupoles should withstand the radiation resulting from 400 fb$^{-1}$ to 700 fb$^{-1}$, but some nested-type corrector magnets could fail at around 300 fb$^{-1}$. The most likely failure mode is sudden electric breakdown, entailing serious and long repairs. Replacement of the triplet must be anticipated before radiation damage reaches the level where serious failure is a significant possibility.

The replacement can be coupled with an increase in the quadrupole aperture to allow room for an increase in the luminosity via a lower $β^*$. However, larger aperture inner triplet (IT) quadrupoles and the increased luminosity, with consequent higher radiation levels, imply the redesign of the whole interaction region (IR) zone. This redesign includes larger D1 and D2 dipoles, a new electrical feedbox (DFBX), and much better access to various components for maintenance. In addition, larger aperture magnets in the matching sections will be required, as well as a redesign of the collimation system in the high luminosity insertions.

To maximize the benefit of such a long shutdown, this work must be complemented by a series of improvements and upgrades for other systems, and must be coupled with a major upgrade of the experimental detectors. Both the machine and the detectors must be partially redesigned in order to withstand the expected level of integrated luminosity. The upgrade should allow the delivery of 3000 fb$^{-1}$, i.e. one order of magnitude greater than the nominal LHC design goal.

It is clear that the change of the inner triplets in the high luminosity insertions is the cornerstone of the LHC upgrade. The decision for the HL-LHC has been to rely on the success of the advanced Nb$_3$Sn technology, which provides access to magnetic fields well beyond 9 T, allowing the maximization of the aperture of the IT quadrupoles. A 15-year-long study led by the DOE in the US under the auspices of the US LARP programme (see Chapter 1, Section 1.3.2), and more recently by other EU programmes, has shown the feasibility of Nb$_3$Sn accelerator magnets. For the HL-LHC, some 24 IT Nb$_3$Sn quadrupoles are needed: they all feature a 150 mm aperture and an operating gradient of 140 T/m, which entails more than 12 T peak field on the coils. The Q1 and Q3 quadrupoles each consist of a pair of 4 m long magnets, while Q2a and Q2b each consist of a single unit almost 7 m long (see Chapter 3, Section 3.2). The same Nb$_3$Sn technology will be used to provide collimation in the DS, which will be achieved by replacing a number of selected main dipoles with two shorter 11 T Nb$_3$Sn dipoles (see Chapter 11 Section 11.3). A collimator will be installed between the shorter dipoles (see, for example, Ref. [15] and references therein).

In addition to the IT quadrupoles, there are four new D1/D2 separation/recombination dipole pairs, a number of matching section (MS) quadrupoles, not only in IR1 and IR5, but also in IR6, and a smaller number of lattice sextupoles that can be made using well-known Nb-Ti technology (see Tables 2-3 and 2-4). These magnets will feature a larger aperture and will be exposed to higher radiation doses if not properly protected, and thus will be more challenging than the present LHC equivalents (see Chapter 3).

The corrector packages in the IT and in the MS regions need to be significantly upgraded to increase aperture and (where needed) strength. Some 70 corrector magnets of various orders (from dipole for orbit correction to dodecapole skew correctors) and typology (from superferric to nested cos theta) have to be installed with the new larger IR magnets.

### 2.2.2 TAXS/TAXN absorbers

The change of the IT aperture will require replacement of the TAS, the first absorber on either side of the high luminosity interaction points. The TAS protects the downstream magnets from collision debris. Its aperture roughly scales with the IT aperture. The new absorber, named TAXS, will have an aperture of 54 mm (compared with 30 mm in the present TAS), and will have to withstand a flux of particles five times larger than in the present nominal design. In the current LHC, the TAS is probably the most highly activated component of the whole machine. The baseline choice at present is to replace the TAS with the TAXS during LS3 (see Chapter 8).

Given the fact that the experimental detectors have reduced the size of the vacuum chamber by nearly 50% (from 55 mm down to about 35 mm) it is clear that all challenges at the machine–detector interface are increased. This includes keeping background radiation in the detectors at acceptable levels.

### 2.2.3 Crab cavities

Superconducting (SC) RF crab cavities (CC) in the HL-LHC are needed in order to compensate for the geometric reduction factor, thus making the very low $\beta^*$ fully useful for luminosity. HL-LHC crab cavities are beyond the state-of-the-art in terms of their unconventional, compact design, which cannot be achieved with the well-known geometry of an elliptical cavity. They also demand a very precise control of the phase of the RF (to better than 0.001°) so that the beam rotation given before collision is exactly cancelled on the other side of the interaction point (IP). The crab cavities will also pose new challenges for machine protection. Compact crab cavities will be installed on both sides of IP1 and IP5 without additional magnetic doglegs (as in IP4 for the accelerating cavities). Each cavity is designed to provide a transverse kick voltage of 3.4 MV. There are four crab cavities per beam on each side of the IP. They will be assembled in cryomodules, each containing two cavities. All four cavities may be used to rotate the beam in the crossing plane; alternatively, a single cryomodule (two cavities) can be used for this task, with the cavities in the second cryomodule providing a deflection in the orthogonal plane, enabling the so-called crab kissing scheme for reducing the pile-up density [16]. At present, the baseline is to use all crab cavities for geometric compensation, i.e. rotation in the crossing plane.

The first-generation, proof-of-principle, compact crab cavities have recently been tested successfully (see Chapter 4). However, a second generation with machine-oriented characteristics are now under construction by LARP, CERN, and UK institutions (Lancaster University, STFC, and the Cockcroft Institute). A full cryomodule will be tested in the SPS before LS2, to investigate experimentally the effect on a proton beam and to gain the necessary experience in view of LHC operation.

### 2.2.4    Collimation

The collimation system has been designed for the first phase of LHC operation. It is currently operating according to design. However, the impedance of the collimation system may need to be reduced if beam instabilities are triggered at intensities close to, or just above, nominal. Hints of this behaviour have been already seen during Run 1: only operation near nominal conditions can dismiss or validate this hypothesis.

Safe handling of a beam of 1 A or more, with beta functions at collision beyond the design value, will constitute new territory. The triplet must remain protected during the large change of the collision beam parameters ($\beta^*$ transition from 6 m to 10–15 cm). This will be one of the most critical phases of HL-LHC operation: just the beam halo itself could be beyond the damage limit. An upgrade of the collimation system is thus required. The main additional needs associated with the upgrade are a better precision in alignment and materials capable of withstanding higher power.

A second area that will require special attention in connection with the collimation system is the dispersion suppressor (DS), where leakage of off-momentum particles into the first and second main superconducting dipoles has already been identified as a possible LHC performance limitation. The most promising concept is to substitute an LHC main dipole with a dipole of equal bending strength (120 T·m) obtained by a higher field (11 T) and shorter magnetic length (11 m) than those of the LHC dipoles (8.3 T and 14.2 m). The space gained is sufficient for the installation of special collimators. This system is already needed for Run 3 ion operation in the DS region around IP2 following the upgrade of ALICE in LS2. It might also be needed in the DS around IP7 for HL-LHC operation. The requirements in other insertion regions have yet to be assessed.

### 2.2.5    New cold powering

While a considerable effort is under way to study how to replace the radiation-sensitive electronics boards with radiation-hard cards, another solution is also being pursued for special zones: removal of the power converters and electrical feedboxes (DFBs), delicate equipment associated with the continuous cryostat, out of the tunnel. Besides improving LHC availability (fewer interruptions, faster interventions without the need for tunnel access), radiation dose to personnel would be reduced as well.

Removal of power converters and DFBs to locations far from the beam line, and possibly to the surface, is only possible through the use of a novel technology: superconducting links (SCLs) made out of high-temperature superconductors (YBCO or Bi-2223) or $MgB_2$ superconductors. Regions where this radical solution will be needed because of the high radiation load on electronics, and/or the 'as low as reasonably achievable' principle (ALARA), have been identified.

- The long straight section of IP7 where a 500 m cable rated at 20 kA is needed.
- The high luminosity insertion regions IR1 and IR5, where much higher current cables (150 kA and 164 kA) are needed for the IT magnets and the magnets in the MS region (i.e. from D2 to Q6). In this latter case, the superconducting cable will link the magnets with power converters on the surface, with significant challenges to the cryogenics and system integration resulting from the 100 m or so difference in altitude.

### 2.2.6  Enhanced machine protection and remote handling

Various systems will become a bottleneck with aging of the machine and higher performance beyond the 40 fb$^{-1}$ to 60 fb$^{-1}$ per year envisaged in the original LHC design. One such system is the quench protection system (QPS) of the superconducting magnets. The QPS should:

i)  become fully redundant in case of power loss;

ii) allow low energy discharge on quench heaters and easy adaption of the detection thresholds;

iii) provide an interlock for the quench heater discharge based on a sensor for quench heater integrity.

In general, the QPS will need a complete renovation after 2020.

Machine protection will have to be improved, and not just because of the higher beam energy and energy density: it will have to cope with very fast events generated, for example, by crab cavities and by a possible increase of the events generated by falling particles (UFOs).

The LHC has not been designed specifically for remote handling. However, the level of activation from 2020, and even earlier, requires careful study and development of special equipment to allow replacement of collimators, magnets, vacuum components, etc. according to the ALARA principle. While full robotics are difficult to implement given the conditions, remote manipulation, enhanced reality, and supervision are the key to minimizing the radiation dose to personnel.

### 2.2.7  New cryogenics plants and distribution

To increase the flexibility for intervention and rapid restoration of availability (i.e. to minimize loss of integrated luminosity) it will be useful to install a new cryogenics plant in P4 for a full separation between superconducting RF and magnet cooling. This should be done during LS2, to avoid a possible weak zone for Run 3. The new plant should also be able to provide cooling to new cryogenic equipment under consideration for IP4, i.e. a new SCRF harmonic system and the hollow e-lens for halo control, which requires a superconducting solenoid. For the time being, these two systems are not in the baseline; however, they constitute interesting options under study.

A further consolidation that is deemed necessary in the long term is the separation between the cooling of the inner triplets and the few stand-alone superconducting magnets in the MS from the magnets of the arc. The present coupling of IR and arc magnets means that an intervention in the triplet region requires warm-up of the entire sector (an operation of three months, not without risk).

New power plants will be needed to cope with higher heat deposition from the high luminosity points. In particular, given the luminosity-driven heat load in the cold magnets, and the cooling of superconducting crab cavities at 1.9 K, the power (at 4.2 K) of the new cryo-plant in IP1 and IP5 will have to be in the 15–18 kW range. The cooling scheme includes separation, with possible interconnection, between arc and IR cryogenics to gain in flexibility.

### 2.2.8  Enhanced beam instrumentation

Improving beam instrumentation is a continuous task during routine operation of an accelerator. The HL-LHC will require improved or new equipment to monitor and act on proton beams with more challenging parameters than those of the LHC. A short illustrative list includes the following.

- New beam loss monitors for the IT quadrupoles.

- A radiation-tolerant Application-Specific Integrated Circuit (ASIC) for the beam loss monitoring system.

- A new beam position monitoring system, including a high-resolution orbit measurement system, and high-directivity strip-line pick-ups for the insertion regions.

- Emittance measurement: while improving the present system, a new concept-based beam gas vertex emittance monitor is envisaged for the HL-LHC.

- Halo diagnostics to control the halo in order to minimize losses (and especially loss peaks) in the presence of a beam with a stored energy close to 0.7 GJ. Synchrotron radiation imaging, and possibly wire scanners, appear to be the only candidates for halo monitoring in the HL-LHC.

- Diagnostics for crab cavities: electromagnetic pick-ups and streak cameras are being studied for beam shape monitoring.

- Luminosity measurements with new radiation-hard devices (located in the new TAXN) capable of withstanding the radiation level, which will be ten times higher.

### 2.2.9 Beam transfer and kickers

The higher beam current significantly increases the beam-induced power deposited in many elements, including the injection kicker magnets in the LHC ring. New designs for several components in the dump system devices will probably be needed because of the increased energy deposition in the case of direct impact, and because of an increased radiation background, which could affect the reliability of this key machine protection system.

A non-exhaustive list of the elements that could need an improvement or a more radical upgrade (based on the experience from Run 2) is given below.

- Injector kicker magnets (better cooling of the magnets to cope with beam-induced heating, different type of ferrites with higher critical temperature, coating of ceramic tubes to reduce SEY to suppress e-cloud effects).

- Beam dump block TDE with its $N_2$ overpressure system and window VDWB: if these are not compatible with HL-LHC intensities, extension of the dilution pattern may be the only practical and safe solution, implying the installation of additional dilution kicker systems MKB (up to 50%).

- Injection absorber, auxiliary protection collimators, protection masks.

- Beam dump absorber system.

## 2.3 Baseline optics and layout

### 2.3.1 Basic optics and layout choices for the High Luminosity insertions

The current baseline optics design (HLLHCV1.1) has evolved from the previous LHC Upgrade Phase I project [17–19]. A realistic, cost-efficient and robust (achromatic) implementation of low $\beta^*$ collision optics requires the deployment of the Achromatic Telescopic Squeeze (ATS) scheme, together with the installation of insertion magnets of larger aperture [20–24]. Successful validation tests of the ATS with beam were achieved in 2011–2012 [25–29] in very specific conditions (low intensity, no crossing angle to save aperture, etc.). The corresponding number, type, and specifications of the new magnets to reach low $\beta^*$ [20, 21] were then endorsed by the project (see, for example, [30] and references therein).

The historical development of the optics design is summarized in Ref. [31]; here, only the last three parts of this long chain are mentioned, namely the so-called SLHCV3.1b [32], HLLHCV1.0 [33], and HLLHCV1.1 (current baseline) optics. SLHCV3.1b uses ATS optics based on 150 T/m $Nb_3Sn$ triplets and displacement of D2 for crab cavity integration [32]. HLLHCV1.0 is similar to SLHCV3.1b, but with a new triplet layout based on 140 T/m $Nb_3Sn$ triplets [33]. HLLHCV1.1, the new baseline, is based on HLLHCV1.0, but with some modifications to take into account the results of design studies for D2, energy deposition studies for the passive protection of the superconducting elements, hardware integration studies, and updated naming conventions [34, 35], and corresponding optical configurations.

Table 2-2 presents an overview of the main features of the three layouts and of the corresponding optical configurations.

Table 2-2: Main HL-LHC optics variants currently under study. The baseline collision optics corresponds to $\beta^* = 15$ cm in both transverse planes (round optics) with a full crossing angle of 590 μrad. Other collision optics are available, round or flat, for dedicated studies.

|  | SLHC V3.1b | HLLHCV1.0 | HLLHCV1.1 (Baseline) |
|---|---|---|---|
| Collision $\beta^*$ IP1, IP5 | Round: 15cm, (10 cm, 33 cm, 40 cm). Flat: 30/7.5cm, (20/5 cm) with HV, VH crossing. | Round: 15cm, (10 cm). Flat: 30/7.5cm, (20/5 cm) with HV, VH crossing. Complete squeeze. | |
| Pre-squeeze $\beta^*$ IP1, IP5 | 40 cm, (2 m) | 44 cm, (3 m) transition strengths. | 44 cm |
| Injection $\beta^*$ IP1, IP5 | 5.5 m (11 m) | 6 m, (11 m, 18 m) | 6 m (15 m) |
| Triplet gradient | 150 T/m | 140 T/m | 140 T/m |
| Triplet magnetic length | Q1–Q3: 7.685 m  Q2: 6.577 m | Q1–Q3: 4.002 m × 2  Q2: 6.792 m | Q1–Q3: 4.00 m × 2  Q2: 6.8 m |
| Triplet corrector package | Nested triplet nonlinear corrector package with new $a_5$, $b_5$, $a_6$ corrector coils | Superferric, non-nested, nonlinear corrector package. | |
| Insertion region dipoles | D2 moved towards the IP by 15 m. For version HLLHCV1.1 the magnetic length of D1 [40] and D2 has been shortened. | | |
| Insertion region quadrupoles | MQYY type for Q4 in IR1, IP5. Q5 moved towards arc by 11 m. MQYL type for Q5 in IR1, IR5, IR6. Additional MS in Q10 of IR1 and IR5. | MQYY type for Q4 in IR1, IR5. Q5 moved towards the arc by 11 m. MQYL type for Q5 in IR1, IR5, IR6. Additional MS in Q10 of IR1 and IR5. | MQYY type for Q4 in IR1, IR5. Q5 moved towards arc by 11 m. Q4 moved towards arc by 8 m. MQY at 1.9 K type for Q5 in IR1, IR5. Double MQY for Q5 in IR6. Additional MS in Q10 of IR1 and IR5. |
| Crab cavities |  | 3 | 4 |

The current baseline layout incorporates various optimizations, and in particular has been made compatible with the latest hardware parameters and constraints. The magnetic elements in the region between the IP and Q4 (Figure 2-3) have been positioned to optimize the strength requirements for the magnets and for ancillary equipment. For instance, moving the Q4 quadrupole changes the value of the beta functions at the location of the crab cavities, thus improving their efficiency.

In the triplet region (Figure 2-3, which is in the range of approximately 20 m to 80 m) the Q1 and Q3 magnets are split into two and the dipole corrector magnets (used to create the crossing and separation schemes) are implemented in a nested configuration for both planes. The corrector package close to Q3 consists of superferric magnets. The specifications and performance of the non-linear correctors (used to compensate the field quality effects of the triplets and D1 separation dipoles on both sides of the IP) are reported in Refs. [36, 37]. Detailed numerical simulations indicate that additional corrector types are needed to cope with the pushed performance of the HL-LHC, so the layout of the correctors will not be a simple carbon copy of the existing layout. Inclusion of the field quality of the D2 separation dipole has been considered, but is not trivial, due to the two-in-one structure of the D2.

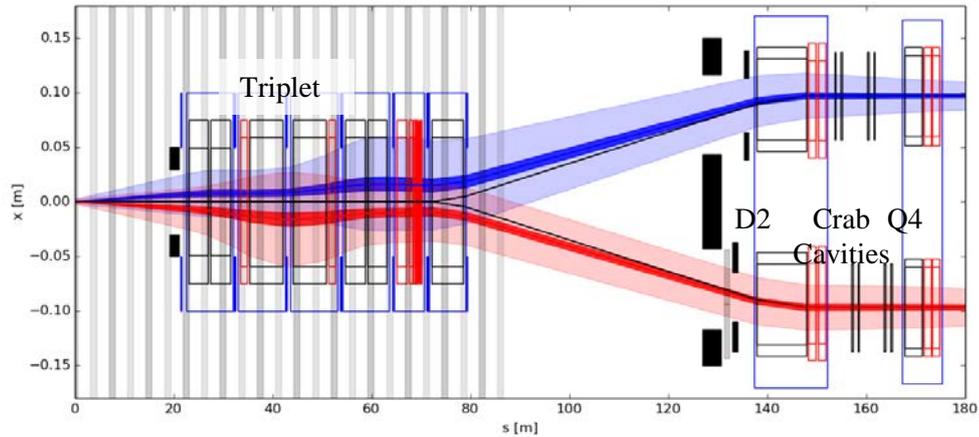

Figure 2-3: Overall layout of the insertion region between the IP and Q4. The dark blue and red areas represent the 2 $\sigma$ beam envelope. The light regions correspond to a 12 $\sigma$ value of the beam envelope for an emittance of 3.5 μm with a tolerance of 20% for beta-beating and 2 mm of closed orbit distortion. The shaded grey areas in the triplet region represent the locations of the parasitic beam–beam encounters.

The block of two separation dipoles has been changed with respect to the nominal LHC layout, decreasing their separation. The D2 area is particularly delicate for several reasons. First, there are space constraints because of the need for protection devices such as the absorber for neutral debris from the collisions. Second, the transverse separation is not yet optimal, leading to a reduction in the amount of iron between the two apertures of the D2, as well as to reduced beam and mechanical apertures because of the large values of the beta functions at this point. Downstream of D2, the situation is not much easier, as the crab cavities impose tight constraints on the space between D2 and Q4, as well as on the values of the beta functions.

Detailed work has been performed to specify the strengths of dipole orbit correctors in the triplets and the D2 and Q4 magnets [38, 39]. Initially, the required strength needed too long correctors, mainly because of the need to close the bumps at the D2 dipole in order to avoid a non-zero closed orbit at the location of the crab cavities. Detailed analysis of the RF aspects allowed the tight constraints to be relaxed (0.5 mm of closed orbit distortion can be tolerated at the crab cavities when operating and 2–3 mm when the cavities are made transparent to the beam). The design now includes a correction scheme with magnets that are 1.5–2.5 m long. This makes it possible to close the orbit bumps further downstream from the D2 separation dipoles, thus reducing their overall strength. In the current layout, Q4 is a new magnet with a larger aperture, MQYY, while Q5 is a MQY-type (the Q4 of the present LHC) operating at 1.9 K to provide the required gradient.

The implementation of the ATS scheme in the HLLHCV1.0 requires hardware changes in other parts of the LHC ring. In particular, an additional lattice sextupole (MS) magnet should be installed in Q10 in IR1 and IR5. Moreover, Q5 in IR6 should be upgraded. The current baseline layout envisages the installation of a second MQY-type quadrupole close to each existing Q5.

Table 2-3 lists the key parameters of the quadrupoles (new or refurbished) to be installed in IR1 and IR5, while Table 2-4 gives the corresponding parameters for the separation dipoles and orbit correctors. Table 2-5 gives the parameters for the multipolar correctors.

Table 2-3: New or refurbished quadrupoles for HL-LHC, all operating at 1.9 K. The orientation of the rectellipse aperture ($R$) [40] can be changed to optimize the mechanical aperture.

|  | Inner triplet (single aperture) | | | Matching section (two-in-one) | | |
|---|---|---|---|---|---|---|
| Magnet | Q1 | Q2 | Q3 | Q4 | Q5 | Q6 |
| Number | 2 | 2 | 2 | 1 | 1 | 1 |
| Type | MQXFA | MQXFB | MQXFA | MQYY | MQY | MQML |
| Magnetic length [m] | 4.0 | 6.8 | 4.0 | 3.8 | 3.4 | 4.8 |
| Gradient [T/m] | 140 | 140 | 140 | 115 | 200 | 200 |
| Coil aperture [mm] | 150 | 150 | 150 | 90 | 70 | 56 |
| Aperture separation [mm] | - | - | - | 194 | 194 | 194 |
| Beam screen (BS) shape | Octagon | Octagon | Octagon | Rectellipse | Rectellipse | Rectellipse |
| BS aperture (H/V) [mm] | 98/98 | 118/118 | 118/118 | 64/74 | 44/57.8 | 35.3/45.1 |
| Mechanical tolerances (R/H/V) [mm] [41] | 0.6/1/1 | 0.6/1/1 | 0.6/1/1 | 0.84/1.26/0.6 | 0.84/1.26/0.6 | As built |

Table 2-4: New dipole magnets for HL-LHC, all operating at 1.9 K. The orientation of the rectellipse ($R$) aperture can be changed to optimize the mechanical aperture. The orbit correctors can be nested or consecutive as indicated.

|  | Separation/recombination dipoles | | Orbit correctors | | | |
|---|---|---|---|---|---|---|
| Assembly | D1 | D2 | Corrector package | Q2 | D2 | Q4 |
| Number per side per insertion | 1 | 1 | 1 [HV nested] | 2 [HV nested] | 2 [HV consec.] | 2 [HV consec.] |
| Type | MBXF | MBRD | MCBXFA | MCBXFB | MCBRD | MCBYY |
| Magnetic length [m] | 6.27 | 7.78 | 2.2 | 1.2 | 1.5 | 1.5 |
| Integrated field [T m] | 35 | 35 | 4.5 | 2.5 | 4.5 | 4.5 |
| Coil aperture [mm] | 150 | 105 | 150 | 150 | 100 | 100 |
| Aperture separation [mm] | n/a | 188 | - | - | 194 | 194 |
| BS shape | Octagon | Octagon | Octagon | Octagon | Octagon | Rectellipse |
| BS aperture (H/V) [mm] | 118/118 | 84/84 | 118/118 | 118/118 | 79/79 | 64/74 |
| Mechanical tolerances (R/H/V) [mm] | 0.6/1/1 | 0.84/1.36/1 | 0.6/1/1 | 0.6/1/1 | 0.84/1.36/1 | 0.84/1.26/0.6 |

Table 2-5: New multipolar superferric correctors for HL-LHC, all operating at 1.9 K.

| **Number** | 1 | 1 | 1 | 1 | 1 | 1 | 1 | 1 | 1 |
|---|---|---|---|---|---|---|---|---|---|
| Number of poles | 4 | 12 | 12 | 10 | 10 | 8 | 8 | 6 | 6 |
| Normal/skew | Skew | Normal | Skew | Normal | Skew | Normal | Skew | Normal | Skew |
| Type | MQSXF | MCTXF | MCTSXF | MCDXF | MCDSXF | MCOXF | MCOSXF | MCSXF | MCSSXF |
| Magnetic length [m] | 0.807 | 0.43 | 0.089 | 0.095 | 0.095 | 0.087 | 0.087 | 0.111 | 0.111 |
| Integrated field [mT·m] at 50 mm | 1000 | 86 | 17 | 25 | 25 | 46 | 46 | 63 | 63 |
| Coil aperture [mm] | 150 | 150 | 150 | 150 | 150 | 150 | 150 | 150 | 150 |
| BS shape | Octagon | Octagon | Octagon | Octagon | Octagon | Octagon | Octagon | Octagon | Octagon |
| BS aperture (H/V) [mm] | 118/118 | 118/118 | 118/118 | 118/118 | 118/118 | 118/118 | 118/118 | 118/118 | 118/118 |
| Mechanical tolerances (R/H/V) [mm] | 0.6/1/1 | 0.6/1/1 | 0.6/1/1 | 0.6/1/1 | 0.6/1/1 | 0.6/1/1 | 0.6/1/1 | 0.6/1/1 | 0.6/1/1 |

As already mentioned, protection devices are required for the new layout of the IR1 and IR5 regions. The current LHC layout has only a TAS in front of Q1, to protect this magnet from collision debris, and a TAN to protect D2 from the neutrals produced at the IP. For the HL-LHC, these two devices will have to be upgraded to withstand much larger luminosities. Furthermore, additional masks are envisaged to protect other magnets in the matching section. A summary of the characteristics of these devices can be found in Table 2-6.

Table 2-6: New absorbers for HL-LHC, all operating at 1.9 K. The orientation of the rectellipse aperture can be changed to optimize the mechanical aperture.

| | Inner triplet (single aperture) | Matching section (two-in-one) | | |
|---|---|---|---|---|
| Absorber | TAS | TAN | Mask D2 | Mask Q5 | Mask Q6 |
| Aperture | 1 | 2 | 2 | 2 | 2 |
| Type | TAXS | TAXN | TCLMA | TCLMB | TCLMC |
| $L$ [m] | 1.8 | 3.5 | 0.5 | 1.0 | 1.0 |
| Aperture separation [mm] | n/a | 149–159 | 188 | 194 | 194 |
| Aperture (H/V) [mm] | 54/54 | 80/80 | 84/84 | 44/57.8 | 35.3/45.1 |
| Mechanical tolerances (R/H/V) [mm] | 2/0.5/0.5 | 0.6/1/1 | 0.6/1/1 | 0.6/1/1 | 0.6/1/1 |

Figure 2-4 shows example optics configurations for injection and collision. Several configurations can be provided apart from the nominal (i.e. round) optics.

Table 2-7 gives the main sets of $\beta^*$ values (including the optical parameters corresponding to the ion runs). Since IR2 and IR8 are running with increased strength of the triplets at injection, a so-called pre-squeeze has to be applied at top energy to reduce the strength of the triplets at constant value of beta function at the IP.

The transition between the various optical configurations has been studied in detail [42, 43]. The sequence of gradients during the squeeze is available, and will be used to perform first estimates of the hysteresis effects. Moreover, it will be possible to evaluate the time required to accomplish the squeeze, which is essential information to determine specifications for the required power converters. Work is in progress to address these two points; results are expected in the coming months.

Finally, it is worth mentioning additional studies that have looked at alternative layouts. Options have been studied based on triplets using 120 T/m and 170 T/m gradients [44, 45], and an additional Q7 for crab cavity kick enhancements [46] without upgrading the matching section layout [47]. The latest results can be found in Ref. [48].

There are numerous constraints on the layout of components, arising from various considerations. The constraints and the associated issues are described in Chapter 15.

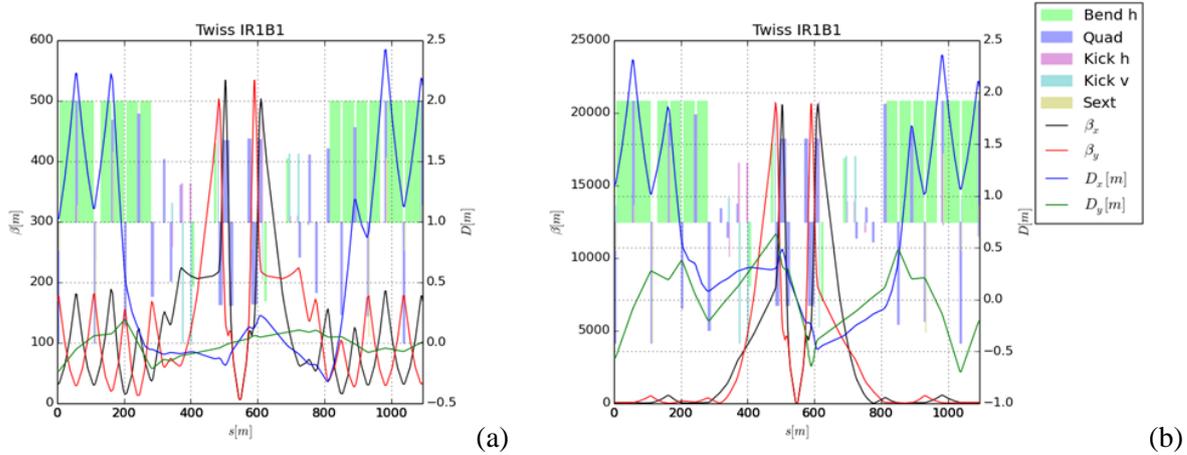

(a)    (b)

Figure 2-4: Optical functions at (a) injection and (b) collision. Note the different vertical scales: at collision, the beta functions in the triplets are large, to provide the low $\beta^*$ at the IP.

Table 2-7: Available optical configurations for the baseline layout. IR3 and IR7 are not reported here as they have static optics from injection to collision and do not take part in the ATS scheme. Some alternative configurations (other than the nominal) are also shown.

| Optics | IR1 | IR5 | IR2 | IR8 | IR4 | IR6 |
|---|---|---|---|---|---|---|
| Injection | $\beta^*$ = 6 m, inj. | $\beta^*$ = 6 m, inj. | $\beta^*$ = 10 m, inj. | $\beta^*$ = 10 m, inj. | Inj. | Inj. |
| End of ramp | $\beta^*$ = 6 m | $\beta^*$ = 6 m | $\beta^*$ = 10 m | $\beta^*$ = 10 m | Inj. | Inj. |
| Pre-squeeze | $\beta^*$ = 44 cm | $\beta^*$ = 44 cm | $\beta^*$ = 10 m | $\beta^*$ = 3 m | Inj. | Inj. |
| Collision round | $\beta^*_{ATS}$ = 15 cm | $\beta^*_{ATS}$ = 15 cm | $\beta^*$ = 10 m, ATS (3×, 3×) | $\beta^*$ = 3 m, ATS (3×, 3×) | ATS (3×, 3×) | ATS (3×, 3×) |
| Collision ions | $\beta^*$ = 44 cm | $\beta^*$ = 44 cm | $\beta^*$ = 50 cm | $\beta^*$ = 50 cm | Inj. | Inj. |
| Collision VDM | $\beta^*$ = 30 m | $\beta^*$ = 30 m | In preparation | In preparation | Inj. | Inj. |
| **Alternative configurations** | | | | | | |
| Collision Flat | $\beta^*_{ATS}$ = 7.5/30 cm | $\beta^*_{ATS}$ = 30/7.5 cm | $\beta^*$ = 10 m, ATS (6×,1.5×) | $\beta^*$ = 3 m, ATS (6×,1.5×) | ATS (1.5×, 6×) | ATS (1.5×, 6×) |
| Collision FlatHV | $\beta^*_{ATS}$ = 30/7.5 cm | $\beta^*_{ATS}$ = 7.5/30 cm | $\beta^*$ = 10 m, ATS (1.5×, 6×) | $\beta^*$ = 3 m, ATS (1.5×, 6×) | ATS (6×,1.5×) | ATS (6×,1.5×) |
| Collision sRound | $\beta^*_{ATS}$ = 10 cm | $\beta^*_{ATS}$ = 10 cm | $\beta^*$ = 10 m, ATS (4.5×, 4.5×) | $\beta^*$ = 3m, ATS (4.5×, 4.5×) | ATS (4.5×, 4.5×) | ATS (4.5×, 4.5×) |
| Collision sFlat | $\beta^*_{ATS}$ = 5/20 cm | $\beta^*_{ATS}$ = 20/5 cm | $\beta^*$ = 10 m, ATS (9×, 4.5×) | $\beta^*$ = 3 m, ATS (9×, 4.5×) | ATS (4.5×, 9×) | ATS (4.5×, 9×) |
| Collision sFlatHV | $\beta^*_{ATS}$ = 20/5 cm | $\beta^*_{ATS}$ = 5/20 cm | $\beta^*$ = 10 m, ATS (4.5×, 9×) | $\beta^*$ = 3 m, ATS (4.5×, 9×) | ATS (9×, 4.5×) | ATS (9×, 4.5×) |

### 2.3.2  Target field quality and dynamic aperture

The dynamic aperture (DA) has been used since the initial steps of the design of the LHC [40] to determine the required field quality of the various magnet classes. For computation of the DA in the HL-LHC, particles are tracked over $10^6$ or $10^5$ turns, depending on whether beam–beam effects are included or neglected, respectively. The initial momentum co-ordinate is set to two-thirds of the bucket height ($2.7 \times 10^{-4}$ and $7.5 \times 10^{-4}$ for collision and injection energy, respectively). Sixty implementations of the random components in the magnets, corresponding to sixty realizations of the LHC lattice, are considered in the numerical simulations. Eleven phase space angles have routinely been used (although for special studies up to 59 values have been probed), while thirty particle-pairs per 2 $\sigma$ amplitude step have been used. All these parameters have been specified during the design stage of the LHC. Since then, the amount of available computing power has increased, thanks to the increased CPU power of the CERN batch system and because of the use of volunteer-based computing resources [49]: this has enabled an increase of the number of directions considered in the studies, making the DA estimate more accurate. Note that the number of turns and random seeds affects the accuracy of the DA calculation, which is at least 0.1 $\sigma$ in this case.

For reference, the multipole expansion used to describe the magnetic field is given as [40]:

$$B_y + iB_x = B_{\text{ref}} \sum_{n=1}^{\infty} (b_n + ia_n) \left(\frac{x+iy}{r_0}\right)^{n-1}, \qquad (2\text{-}5)$$

where $B_x$, $B_y$, and $B_{\text{ref}}$ are the transverse magnetic field components and the reference field, respectively. The coefficients $a_n$, $b_n$ are the skew and normal field components, and $r_0$ is a reference radius. In the framework of the LHC studies the magnetic errors are split into three components, namely mean (*S*), uncertainty (*U*), and random (*R*), such that a given multipole is obtained by:

$$b_n = b_{n_S} + \frac{\xi_U}{1.5} b_{n_U} + \xi_R b_{n_R}, \qquad (2\text{-}6)$$

where $\xi_U$, $\xi_R$ are Gaussian-distributed random variables cut at 1.5 $\sigma$ and 3 $\sigma$, respectively. The $\xi_U$ variable is the same for all magnets of a given class, but changes from seed to seed and for the different multipoles. On the other hand, $\xi_R$ also changes from magnet to magnet.

The target value of the DA differs between injection and collision energies. At injection, where the beam–beam effects can be safely neglected, the focus is on the impact of magnetic field quality. For the LHC design [40], a target value of 12 $\sigma$ (for a normalized emittance of 3.75 μm) was assumed. The best model of the LHC, including the measured field quality of the magnets and the sorting of magnets, provides a DA slightly lower than 11 $\sigma$ [50]. No signs of issues due to DA limitations have been observed during operation or dedicated studies in Run 1, although operation at high intensity has been conducted with beams with an emittance smaller than nominal (2–2.5 μm rather than 3.75 μm).

At top energy, beam–beam effects cannot be neglected and the DA has to be evaluated, including both magnetic field imperfections and head-on and long-range beam–beam phenomena (see Section 2.4.2). Hence, the approach taken consists of probing the impact on DA of the field quality of the new magnets, asking that all new magnets have an impact on the DA that is in the shadow of the triplet quadrupoles. Eventually, the beam–beam effects are also included, providing the final DA value.

Studies for the field quality of the new magnets started from the top energy configuration and with an earlier version of the layout, namely the so-called SLHCV3.1b [32]. This allowed first estimates of the required field quality to be derived, which were then improved by including consideration of the injection energy, where the beam size reaches its maximum and the field quality is worse, due to the persistent current effect. The newer layout HLLHCV1.0 [33] has been used following its release.

In the numerical simulations consideration is made of the machine as built, i.e. the best knowledge of the measured magnetic errors is assigned to the magnets as installed, while, for the magnets that will be replaced according to the upgrade plans, the expected error table, with statistical assignment of errors, is used. This is the baseline configuration of the LHC ring to which magnetic field errors of other classes of magnets can be selectively added.

In these studies the acceptable minimum DA was set to about 10.5 $\sigma$ at top energy, based on experience from the LHC. The DA calculation was performed using long-term tracking in SixTrack [51, 52], neglecting beam–beam effects. Determination of the required field quality based on DA computations is intrinsically a non-linear problem. The field quality obtained from electromagnetic simulations is used as an initial guess. Then, optimization of the field quality essentially involves determining the Jacobian of the DA as a function of the multipoles around the initial value of field quality. For this reason, it is of paramount importance to have a reliable estimate of the expected field quality from detailed electromagnetic simulations and measurements (see Chapter 3). The resulting error tables can be found in the official optics repositories [53, 54] and are also collected in Ref. [55].

The previous IT specifications at 7 TeV [56] were updated to take into account the additional IT correctors for $a_5$, $b_5$, $a_6$ errors. It is worth mentioning that in all the studies reported in this document, the IT correctors have been considered as ideal devices, i.e. exactly correcting the field quality of the IT and D1 magnets, without any error due to uncertainty in the transfer function of the correctors. These specifications will be referred to as IT_errortable_v66. An estimate of the D1 field quality is based on magnet design and referred to as D1_errortable_v1 [57]. Due to the evolution of the D2 dipole design, three versions of the D2 field quality were used in the study: these are referred to as D2_errortable_v3, _v4 [58], and _v5 [59]. The D2 low-order terms at 7 TeV are shown in Table 2-8. Estimates for the Q4 and Q5 magnets are based on a scaling of the measured field of the existing MQY quadrupole and referred to as Q4_errortable_v1 and Q5_errortable_v0, respectively.

Table 2-8: Evolution of low order terms of the estimated D2 field quality at 7 TeV ($r_0 = 35$ mm).

| n | $a_{nS}$ | $a_{nU}$ | $a_{nR}$ | $b_{nS}$ | $b_{nU}$ | $b_{nR}$ |
|---|---|---|---|---|---|---|
| D2_errortable_v3 | | | | | | |
| 2 | 0.0 | 0.679 | 0.679 | 65.0 | 3.000 | 3.000 |
| 3 | 0.0 | 0.282 | 0.282 | -30.0 | 5.000 | 5.000 |
| 4 | 0.0 | 0.444 | 0.444 | 25.0 | 1.000 | 1.000 |
| 5 | 0.0 | 0.152 | 0.152 | -4.0 | 1.000 | 1.000 |
| 6 | 0.0 | 0.176 | 0.176 | 0.0 | 0.060 | 0.060 |
| D2_errortable_v4 | | | | | | |
| 2 | 0.0 | 0.679 | 0.679 | 25.0 | 2.500 | 2.500 |
| 3 | 0.0 | 0.282 | 0.282 | 3.0 | 1.500 | 1.500 |
| 4 | 0.0 | 0.444 | 0.444 | 2.0 | 0.200 | 0.200 |
| 5 | 0.0 | 0.152 | 0.152 | -1.0 | 0.500 | 0.500 |
| 6 | 0.0 | 0.176 | 0.176 | 0.0 | 0.060 | 0.060 |
| D2_errortable_v5 | | | | | | |
| 2 | 0.0 | 0.679 | 0.679 | 1.0 | 1.000 | 1.000 |
| 3 | 0.0 | 0.282 | 0.282 | 1.0 | 1.667 | 1.667 |
| 4 | 0.0 | 0.444 | 0.444 | -3.0 | 0.600 | 0.600 |
| 5 | 0.0 | 0.152 | 0.152 | -1.0 | 0.500 | 0.500 |
| 6 | 0.0 | 0.176 | 0.176 | 2.0 | 0.060 | 0.060 |

The SLHCV3.1b collision optics features $\beta^* = 15$ cm. The desired minimum DA (among all seeds and phase angles) at collision energy is about 10 $\sigma$. Tracking simulations performed with the error table IT_errortable_v66 and without the D1, D2, Q4 and Q5 magnetic errors give $DA_{\min} = 10.4\ \sigma$, which is acceptable. Note that $DA_{\min}$ stands for the minimum DA over all seeds and angles, while $DA_{\text{ave}}$ represents the minimum over all angles of the DA averaged over the seeds.

As a next step, the impact on DA of the field quality in the D1, D2, Q4 and Q5 magnets was verified. The Q4 and Q5 estimated magnetic errors produced negligible effect on the DA, hence their field quality is acceptable. Impact of the D1 estimated errors is mostly due to the relatively large allowed multipoles $b_{n_S}$. It is found that the largest DA reduction is caused by $b_{7S}$ and $b_{9S}$. The low-order D1 errors have negligible effect since they are compensated for by the included IT correctors of order $n$ between 3 and 6. To reduce the impact of $b_{7S}$ and $b_{9S}$ while keeping them realistic, it is proposed to reduce them by a factor of 2 (to 0.2 and −0.295, respectively) relative to D1_errortable_v1.

Two versions of the estimated D2 field quality were used for the SLHCV3.1b tracking: D2_errortable_v3 and D2_errortable_v4. These tables were produced during successive iterations of the field quality optimization. The $b_2$ to $b_4$ terms are rather large due to field saturation. These terms showed a strong impact on the DA. The $b_2$ affected the linear optics by increasing $\beta^*$, thus resulting in a too optimistic DA value. To avoid this effect, the $b_2$ was set to zero for subsequent tracking campaigns, on the assumption that it can be reduced by appropriate design and that $\beta^*$ will be corrected after measurements. It is found that the D2 $b_3$ and $b_4$ have a strong effect on the DA. Effects of feed-down due to the orbit in the straight D1 and D2 magnets were found to be very small. To maintain the $DA_{min}$ close to 10 $\sigma$, $b_3$ and $b_4$ must be reduced by an order of magnitude relative to D2_errortable_v3. These terms had been, indeed, much improved in the updated error table, D2_errortable_v4 (see Table 2-8). Following the tracking results, the proposed further adjustment for D2_errortable_v4 is to reduce $b_2$ to about 1 unit and $b_{3S}$ from 3.0 to 1.5. The resulting $DA_{min}$ and $DA_{ave}$ at 7 TeV, with all new magnet errors and adjustments, are 9.90 $\sigma$ and 11.64 $\sigma$, respectively, which is still acceptable.

The $\beta^*$ for SLHCV3.1b lattice at 450 GeV is 5.5 m, with peak beta functions in the IR magnets lower by a factor of 35 than those in the collision optics. Beam sizes in these magnets are also reduced even though the emittance is larger by a factor of 16. Therefore, the impact of field errors of the new magnets will be much smaller, and the use of the IT correctors is a priori not needed: these results are confirmed by tracking studies. Hence, their field quality at injection based on the present estimates is acceptable. The resulting $DA_{min}$ and $DA_{ave}$ with all errors are 10.16 $\sigma$ and 10.5 $\sigma$, respectively, and are also acceptable. Another option of the injection optics, with $\beta^*$ of 11 m, was verified and showed very similar DA.

The injection DA, however, is about 1 $\sigma$ smaller than the DA of the nominal LHC. Since it is not limited by the IR magnets, other improvements (e.g. in the arcs) may need to be considered. Possible options include a larger integer tune split and adjustment of the working point. Tune scans indicate an effect of the 7th order horizontal resonance close to the current tune (62.28, 60.31). Reducing the horizontal and vertical tunes by about 0.01 would increase the DA by about 0.5 $\sigma$.

HLLHCV1.0 is the latest version of the HL-LHC lattice that has been considered in numerical simulations so far. Some of the differences relative to the SLHCV3.1b include: a longer IT with a lower gradient of 140 T/m and higher peak beta function (7%) at collision, adjusted orientation of magnets in the cryostat, new IT corrector positions, and different phase advance between IP1 and IP5. Using the previously optimized field quality of the new magnets, the collision DA of the HLLHCV1.0 lattice is reduced by about 1 $\sigma$ relative to the SLHCV3.1b, i.e. with $DA_{min}$ and $DA_{ave}$ of 8.8 $\sigma$ and 10.4 $\sigma$, respectively. A stronger impact of the D2 $b_3$ and $b_4$ terms of the previously adjusted D2_errortable_v4 was noticed. Since $b_{3S}$ had been already reduced in this table, the next step was to reduce $b_{4S}$ by half. This improved the DA to $DA_{min}$ = 9.1 $\sigma$ and $DA_{ave}$ = 11.1 $\sigma$. Further improvement was achieved when using the most recent D2 field estimate D2_errortable_v5 [59] (see Table 2-8), where the $b_2$ and $b_3$ terms are reduced at the expense of somewhat larger higher-order terms. In this case, $DA_{min}$ = 9.8 $\sigma$ and $DA_{ave}$ = 12.5 $\sigma$, as shown in Figure 2-5(a), which is acceptable and rather comparable to the DA of the SLHCV3.1b. The reasons for such noticeable improvement will need to be further analyzed.

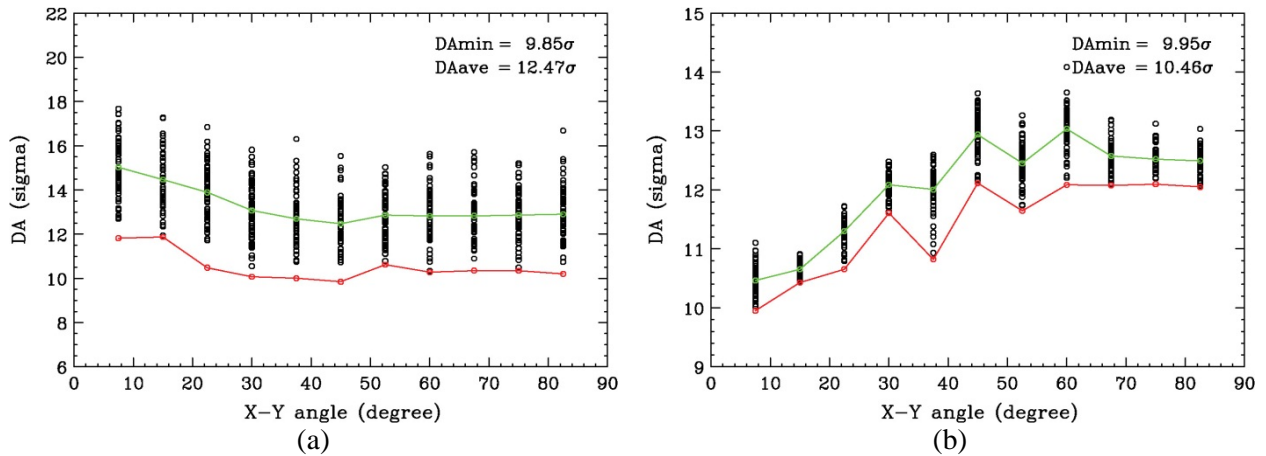

Figure 2-5: DA of HLHCV1.0. (a) DA at 7 TeV with adjusted estimated field quality of new magnets. (b) DA at 450 GeV with estimated field quality of new magnets. The r.m.s. beam size is that corresponding to a normalized emittance of 3.75 μm.

The $\beta^*$ at injection for HLLHCV1.0 is 6.0 m, comparable to SLHCV3.1b. The impact of the field errors in the new magnets on the DA was verified and found to be insignificant (similar to SLHCV3.1b). The resulting DA with all errors ($DA_{min}$ = 9.9 $\sigma$ and $DA_{ave}$ = 10.5 $\sigma$) is acceptable, see Figure 2-5(b).

Field quality and dynamic aperture studies will be pursued in the future along several lines, including:

- dedicated studies to assess the impact of field quality of IT, D1, D2, Q4 and Q5 on linear optics, knowing that the distortion of the optical parameters can stem from both the $b_2$ component and the feed-down from $b_3$ and $a_3$ via the crossing scheme bumps;
- dedicated studies (ongoing) to assess the maximum tolerable ripple in the power converters of the IT quadrupoles and magnets in the matching section [60];
- specification of crab cavity field quality: preliminary results [61–64] seem to indicate that the estimated field quality should be good enough to prevent any impact on the DA;
- assessment of the impact of fringe fields for the large aperture magnets, including the new IT quadrupoles and the separation dipoles.

Regarding fringe fields, the quadrupolar component has already been considered and found non-problematic in Ref. [65]. Preliminary analytical results [66] indicate that, albeit small, the detuning with amplitude induced by the fringe fields is not completely negligible, thus calling for a second level of study. This should include long-term numerical simulations to study the non-linear effects generated. This opens the wide field of symplectic integration as, in the presence of 3D magnetic fields, the standard approach based on multipoles cannot be applied. Work is underway to study the best integration schemes and their implementation [67, 68], before starting the real numerical work.

## 2.4 Performance

### 2.4.1 Beam stability

The impedance in the HL-LHC, shown in Figure 2-6 [69], is not dramatically higher than in the LHC. Molybdenum-coated secondary collimators could decrease the total impedance by more than a factor of 2. However, special caution should be given to devices in high beta regions and unshielded elements.

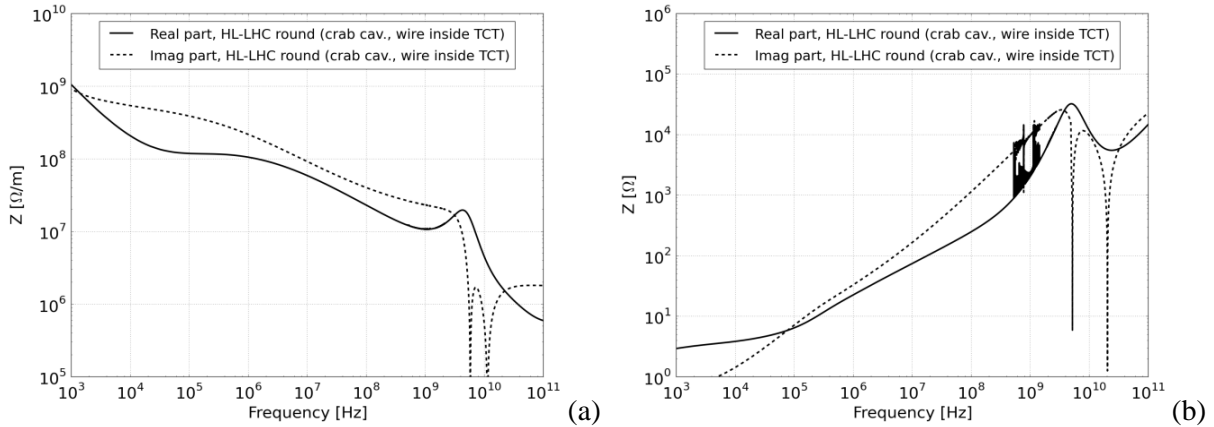

Figure 2-6: (a) First estimate of the horizontal dipolar impedance of HL-LHC, including both crab cavities and beam–beam wire compensators. The vertical dipolar impedance is similar to the horizontal. (b) Longitudinal impedance.

Longitudinal instabilities are not expected to be an issue in the HL-LHC [70]. Single bunch measurements in the LHC at 4 TeV showed an intensity threshold at $1 \times 10^{11}$ p/b, for an RF voltage of 12 MV and a longitudinal emittance of 1 eVs (4 $\sigma$ bunch length of 0.8 ns, scaled from the measurement of the full width at half maximum). Scaling to HL-LHC parameters (16 MV, 2.5 eVs) leads to an intensity threshold of ~3.4 × $10^{11}$ p/b. A double RF system is therefore not needed for beam stability in the longitudinal plane. However, a high harmonic RF system [71–77] in bunch shortening mode could provide an additional margin for longitudinal stability; in bunch lengthening mode, by flattening the bunch profile, a high harmonic RF system could reduce intra-beam scattering (IBS) emittance growth rates, beam-induced heating, and pile-up density. A preliminary cavity design for the 800 MHz RF system exists [78]. Recently, the use of a low harmonic RF system in the LHC (200 MHz) as the fundamental RF system has been suggested [79] since:

- it would allow to accept larger longitudinal emittance and therefore larger bunch population from the SPS after its upgrade;
- it could help to reduce IBS, beam-induced heating, and e-cloud effects;
- together with the existing 400 MHz RF system, it could be used for luminosity and pile-up levelling;
- it also has a beneficial effect for ions and the momentum slip-stacking scheme in the SPS [80].

A new design has been proposed for a compact superconducting cavity [81]. The compatibility of this scheme with 400 MHz crab cavities or the possibility of installing 200 MHz crab cavities needs to be studied further if this scenario is to be considered. Finally, the expected benefits of a double RF system should be weighed against the impedance increase and the possible reduced reliability.

Transverse instabilities are a concern based on the experience of the LHC Run 1, during which a transverse instability at the end of the betatron squeeze could not be cured [70]. While transverse mode coupling instabilities (TMCI) thresholds well exceed the nominal HL-LHC bunch population both at injection and at high energy, to achieve transverse single-beam stability the collimators will need to be coated with 5 μm of molybdenum [70], and the transverse damper must be able to damp coupled-bunch instabilities up to the maximum frequency (20 MHz). Figure 2-7 shows the expected HL-LHC single-beam stability limits for different scenarios with a transverse damper. If the instabilities observed in 2012 are mainly single-bunch (and therefore beyond the range of the damper), we will not be able to stabilize the HL-LHC beams in the case of the standard material collimators and RF dipole crab cavities. However, beam stability could be recovered with molybdenum-coated collimators. This assumes that the transverse damper can damp all coupled-bunch modes otherwise we will not be able to stabilize the HL-LHC beams for any scenario. It is thus clear that the operation of the transverse damper is vital for HL-LHC. The situation improves with the negative polarity of the Landau octupoles but the results are qualitatively the same.

In Figure 2-7, stability thresholds are plotted versus transverse emittance, assuming the fully squeezed ATS optics (with 15 cm $\beta^*$ in both planes) [25]. The high beta functions greatly enhance the impedance contribution from crab cavities compared to the un-squeezed flat-top situation. The design of the crab cavities has to be carefully evaluated and optimized with regard to the high-order modes (HOMs), to minimize the contribution to the coupled-bunch instabilities. The coupled-bunch modes will have to be stabilized by the transverse damper; the damper specifications will therefore have to be finalized once the design of the crab cavities is more advanced (at present, the impedance model of the crab cavities is quite coarse and needs to be improved). During the squeeze, the tune spread (providing Landau damping) could be increased if the ATS can be implemented starting from $\beta^* = 2$ m, thanks to the enhanced beta functions in the arcs. This effect would improve very significantly the situation in terms of stability. In reality, the situation will be even less critical because of luminosity levelling, as the smallest $\beta^*$ will be reached only at a lower intensity. All these effects should be studied in more detail in the future.

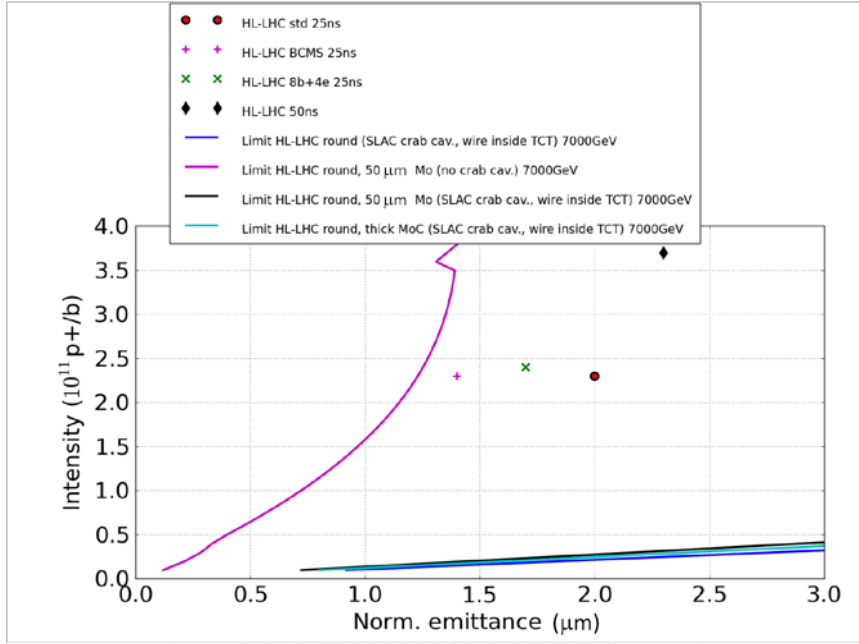

Figure 2-7: Single-beam (25 ns) intensity limit vs. transverse emittance with transverse damper (50-turn damping time) at top energy, for a chromaticity $Q' \sim 15$ for the two extreme cases (CFC collimators with ODU/SLAC crab cavities and Mo-coated collimators without crab cavities) and for positive polarity of the Landau octupoles.

With an additional 800 MHz RF system, the bunch-shortening mode should be preferred from the point of view of transverse beam stability as, even at low chromaticity, beam stability could be reached [70]. However, the operating mode that has been envisaged until now is bunch lengthening to create flat bunches; for that reason alternative scenarios (e.g. bunch flattening by band-limited RF phase noise on the main 400 MHz system) should be studied.

The effect of the electron cloud on beam stability still needs to be assessed. Induced heat loads from the electron cloud are discussed in Section 2.4.3.

Based on experience from LHC Run 1, the interplay between impedance, Landau octupoles, and beam–beam is expected to play an important role in defining the stability limits. A first estimate of the stability diagram for the fully squeezed optics ($\beta^* = 15$ cm) in the presence of both octupoles and long-range beam–beam can be found in Ref. [82]. The stable region with octupoles only increases for $\beta^*$ below 40 cm, and is about 2.5 times larger with the ATS compared to the nominal optics, due to the larger beta functions at the octupoles. As is the case for the LHC, negative octupole polarity (negative amplitude detuning) is preferred for single-beam stability. On the other hand, compensation between negative amplitude detuning from the octupoles and positive amplitude detuning from long-range beam–beam leads to a reduction of the stable region during the squeeze [83]. Below a certain beam–beam separation, positive octupole polarity starts to give larger stable regions and is therefore preferred. Some details of the simulation still need to be checked. However, the proposed operational scenario for the HL-LHC is first to bring the beams into collision, and then to squeeze. By taking advantage in this way as soon as possible of the large amount of Landau damping provided by head-on beam–beam interactions, this should remove possible instability issues arising from long-range beam–beam during the betatron squeeze, while keeping negative octupole polarity (which is better for single-beam stability).

2.4.2   Beam–beam effects

The beam–beam interaction is known to be an important factor limiting the performance reach of present particle colliders. Two of the most significant effects of beam–beam interactions are: (i) the induced particle losses that decrease the beam lifetime, create a high background load for physics experiments, and elevated heat and radiation load on the collimation system; and (ii) the degradation of beam quality manifesting itself through the beam size blow-up that decreases the luminosity delivered to particle physics experiments.

Owing to the extensive theoretical and simulation campaign during the design of the LHC collider, the beam–beam effects in the present machine are well controlled [84]. However, the HL-LHC represents a quantitative as well as a qualitative leap into unknown territory with respect to beam–beam effects. The baseline configuration makes use of some novel concepts that have not so far been used to their full extent in hadron colliders, and thus require careful evaluation. The concepts related to beam–beam effects are: (i) luminosity levelling by variation of the beta function at the IPs; (ii) tilting bunches in the main IPs with the use of RF crab cavities; (iii) significantly high value of the head-on beam–beam tune shift.

Hence, the expected impact of beam–beam interactions on HL-LHC machine performance has been evaluated in order to provide an insight into possible limitations. The studies were mostly performed with the use of the weak–strong approximation and employed the SixTrack and Lifetrack codes, which have been successfully used for the design and optimization of past and existing colliders [85, 86]. Both codes are capable of calculating the area of stable motion in phase space (the dynamic aperture), and hence a direct comparison of the results is possible. The performance reach for weak–strong codes is a few million turns, which is equivalent to a few minutes of machine time. Where necessary, strong–strong simulations with BeamBeam3D, COMBI and a code by K. Ohmi [87–99] were carried out.

In the evaluation of the HL-LHC, the criteria used for establishing satisfactory beam dynamics behaviour were the same as in the LHC design study. In particular, the target value for the one-million turn DA was 6 $\sigma$ (for the nominal HL-LHC emittance of 2.5 μm) or more. The motivation for the choice of such a margin is explained in Ref. [85]. In short, the beam–beam driven diffusion at small amplitudes is quite slow, and the $10^6$ turns of tracking typically does not represent the real long-term stability boundary. In the majority of studies, the 6 $\sigma$ DA corresponds to a true stability boundary of about 4 $\sigma$ with the appearance of chaotic spikes [85]. Benchmarking of the simulations with machine studies seem to indicate that losses and reduction of beam lifetime start to appear only at values of the crossing angle for which the simulated dynamic aperture is as low as 4 $\sigma$ [89, 90]. However, it must be noted that other studies indicate that the simulations of the dynamic aperture of the installed LHC overestimate the dynamic aperture by 20–30% [94].

In the baseline HL-LHC scenario (25 ns spacing), bunches will begin colliding with $2.2 \times 10^{11}$ p/b and transverse normalized emittance of 2.5 μm. The bunches will be tilted by crab cavities at each of the two main IPs to ensure head-on collisions despite the trajectories crossing at an angle. The luminosity will be levelled at the constant value of $5 \times 10^{34}$ cm$^{-2}$ s$^{-1}$ by varying the beta function from ~69 cm at the beginning of the fill to 15 cm at the end, in the case of constant crossing angle of 590 μrad. Assuming negligible transverse emittance growth, the separation of beams at parasitic crossings will thus vary from 26 $\sigma$ at the beginning of the fill to 12.5 $\sigma$ at the end. Hence, from the point of view of beam–beam effects, three stages can be distinguished over the course of a fill.

- Beginning of fill ($\beta^* = 69$ cm, $N = 2.2 \times 10^{11}$ p/b): weak long-range interactions (26 $\sigma$ separation) and strong head-on interactions, characterized by beam–beam tune shift $\xi = 0.031$ (assuming head-on collisions in LHCb), determined by beam brightness $B \sim N/\varepsilon$. Since optics with $\beta^* = 69$ cm were not readily available, simulations were performed with $\beta^* = 40$ cm, which, with the nominal initial intensity, corresponds to a significantly enhanced beam–beam effect (worst-case scenario). The bunch intensity at the specified levelled luminosity is $N = 1.7 \times 10^{11}$ p/b.

- Middle of fill ($\beta^* = 33$ cm, $N = 1.5 \times 10^{11}$ p/b): appreciably large long-range and head-on interactions.

- End of fill ($\beta^* = 15$ cm, $N = 1.1 \times 10^{11}$ p/b): weak head-on ($\xi = 0.015$ due to the particle burn-off in collisions) and relatively strong long-range (12.5 $\sigma$ separation).

All simulations were performed with lattice version SLHCV3.1b. In addition to IR1 and IR5, the beams also collide with a finite angle at IR8 (LHCb), which further enhances the negative impact of the head-on

beam–beam effects on the dynamics. The long-range effects are weaker thanks to the beneficial effects of the $\beta^*$ levelling [92]. Multipole errors in the IR magnets were included in the simulations as specified in Ref. [94]. A parametric study was performed to establish the robustness of the baseline HL-LHC scenario [93] and in order to determine the optimal crossing angle. Figure 2-8 shows the dependence of the minimum DA value on the crossing angle at IP1 and IP5 for different bunch intensities. The dashed lines indicate the minimum target DA and the nominal crossing angle. For the operational cases described above, the DA is always largely above 6 $\sigma$.

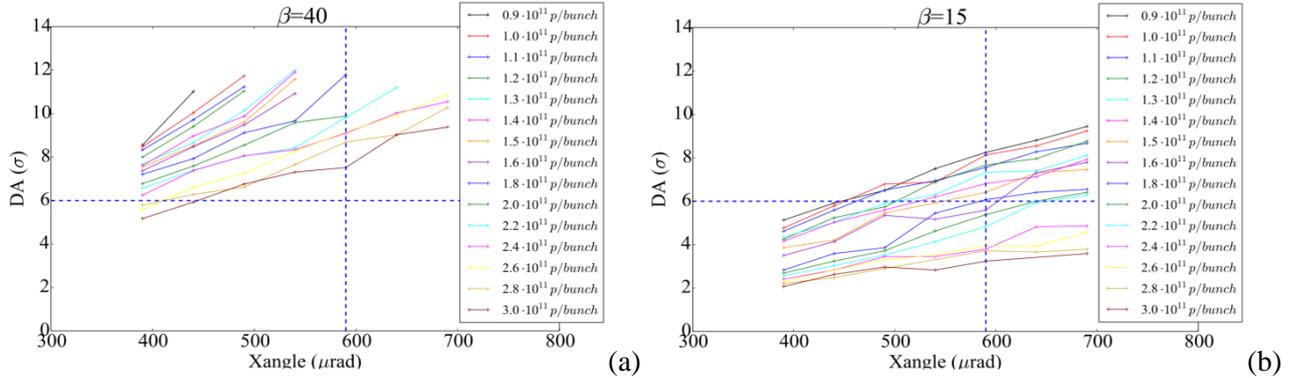

Figure 2-8: Minimum DA for (a) $\beta^* = 40$ cm optics and (b) $\beta^* = 15$ cm optics as a function of crossing angle for different bunch intensities. The r.m.s. beam size corresponds to a normalized emittance of 2.5 µm.

Pacman effects [95, 97] have been evaluated and shown not to have a significant impact on DA and luminosity. The pacman effects are expected to be strongest at the end of the fill ($\beta^* = 15$ cm), and will be weaker than the nominal LHC case due to a larger beam–beam separation (12.5 $\sigma$ compared to 9.5 $\sigma$). With an intensity of $N = 1.1 \times 10^{11}$ p/b and a long-range beam–beam separation of 12.5 $\sigma$ one expects a maximum offset at IP1 and IP5 of about 0.15 $\sigma$. The spread over the bunch train is of the same order, 0.1–0.2 $\sigma$. The long-range variations at IP1 and IP5 result in a very small asymmetry in the tune footprint, and no impact on long-term tracking has been noticed [93].

The results of weak–strong simulations confirm the robustness of the baseline HL-LHC scenario with respect to beam lifetime and particle losses, and suggest that a significant margin exists that would allow either a decrease of the crossing angle to approximately 450 µrad, or operation at higher values of levelled luminosity [95].

Beam–beam effects can induce beam emittance growth and related luminosity lifetime degradation via a variety of mechanisms. Weak–strong simulations of multi-particle bunches were used to evaluate the emittance growth due to beam–beam related betatron resonances. The results predict that the luminosity lifetime due to beam–beam effects will be more than 80 h even in the worst case [98]. A more significant mechanism of emittance degradation can be related to the interplay between the nonlinearity of the beam–beam interaction and various sources of noise. In particular, the phase errors of crab cavities and the ripple of dipole magnet power supplies lead to fluctuations in beam–beam separation. Strong–strong beam–beam simulations have been carried out for the HL-LHC parameters with a large crossing angle and crab cavity compensation [99, 100]. A detailed damper model was included in the simulations. Both $\beta^*$ levelling and crab cavity levelling were simulated including crab cavity noise and dipole power supply ripple [100]. For white random phase noise in the crab cavities, simulations suggest that the r.m.s. noise amplitude should be kept around the level of $3 \times 10^{-5}$ rad in order to maintain a luminosity lifetime of 24 h. This tolerance limit might over-estimate the crab cavity phase noise level since the real phase error will have some spectral distribution different from white noise. For the present studies the spectrum was sampled at a number of frequencies near the betatron frequency. Simulations suggest that strong emittance growth would occur with noise frequencies near the fractional tune of 0.30 and 0.31. The phase errors with those frequencies should be kept as small as possible. The 600 Hz dipole noise was found to have negligible effect on the beam emittance

## 2.4.3 Beam-induced heat load on the cryogenic system

Both impedance and the e-cloud induce heat loads on the cryogenic system [70]. The impedance-induced heat loads with the HL-LHC beam parameters [14] for several key systems are summarized in Table 2-9. In the analysis it is assumed that no forward physics detectors (e.g. ALFA and TOTEM) will be installed during the HL-LHC era. The impedance-induced heat loads for the different types of beam screens vs. temperature are summarized in Table 2-10.

Electron cloud effects should be mitigated by beam-induced scrubbing in the arcs (experience from Run 2 will be vital in that respect) and by low secondary emission yield (SEY) coatings and/or clearing electrodes in the new insertion regions, intended to keep the heat loads within the cooling capacity [70]. Similar actions will be required for the beam screen of the triplets/D1 in IR2 and IR8.

Figure 2-9 shows the heat load from the e-cloud vs. bunch intensity for both an arc dipole and an arc quadrupole, for different SEYs. Provided that a low SEY is achieved, the increased bunch intensity should be acceptable for heat load, but the effect on the beam stability still needs to be assessed. The aim of the scrubbing run is to reach a SEY of ~1.3 in the arc main magnets: this applies for both LHC and HL-LHC, because the dependence of the SEY threshold on bunch intensity in the dipoles is weak. It is worth noting that the quadrupoles have a threshold below 1.1, which cannot be reached by scrubbing. The detrimental effects of the electron cloud in the LHC (heat load in cold regions and emittance blow-up) can be partly mitigated by using specially conceived filling patterns. The underlying idea is to use the flexibility of the injector complex to build bunch trains in LHC with long enough gaps interspersed, to prevent the build-up of an electron cloud along the beam. An alternative scenario (referred to as 8b+4e) based on beams with 25 ns spacing has been conceived to reduce the electron cloud effects in the HL-LHC, if needed, in its initial phase of operation following the upgrade [101] and has been considered as part of the HL-LHC operational scenarios [3]. Operation with a 200 MHz main RF system would allow for longer bunches and would have a positive impact also on the e-cloud [79]. The impact of the bunch length on the e-cloud will be studied experimentally during the coming Run 2 (e.g. to reduce emittance blow-up at low energy).

Table 2-9: Summary of the impedance-induced heat load computations for several key systems

| Element | Expected heat load [W] | Conclusion/comment |
|---|---|---|
| Equipment with RF fingers [102] | Negligible for conforming RF fingers. | Robust mechanical and quality control required during the installation phase. |
| Experimental beam pipes (resonant modes) [103–107] | ATLAS: no significant mode expected.<br>ALICE: potentially* more than 1 kW.<br>CMS: potentially* more than 350 W.<br>LHCb: potentially* more than 250 W. | During Run 2, the temperature should be closely monitored in the large-diameter regions of ALICE, CMS, and LHCb. The impact of these potential expected heat loads on hardware integrity and outgassing should be assessed. |
| All types of beam screens [104] and Table 2-10 | See Table 2-10, where the power losses have been computed vs. temperature (between 20 K and 70 K). | The effects of the beam screen longitudinal weld, the two counter-rotating beams, and the magneto-resistance have been taken into account. Decoupling of the cryogenics for the IR elements and the RF will provide more margin for acceptable heat load in the arcs. |
| Triplet beam position monitors [109] | ~0.2 W/m for the (most critical) 50 ns beam. | This assumes no interferences between the two beams' electromagnetic fields (worst case) and copper coating. |
| New collimators with integrated BPMs and ferrites [110] | ~100 W (of which ~5 W to 7 W would be dissipated in the ferrites, and ~4 W to 6 W in the RF fingers). | More thorough simulation studies as well as bench measurements are under way to confirm these results. This should be acceptable. |
| Injection kickers (MKIs) [111, 112] | Between ~125 W/m and ~191 W/m (based on measurements of 9 MKIs | It is of the order of heat loads estimated with pre-LS1 parameters for the old MKI8D (i.e. before the third |

|  | upgraded to have the full complement of 24 screen conductors). For comparison, most of the MKIs before LS1 had a power deposition of ~ 70 W/m (which did not limit LHC operation). | Technical Stop of 2012) which had a 90° twist in the screen conductors and that had a power deposition of ~160 W/m, based on measurements of impedance during LS1. This required significant time to cool-down after physics fills. For the HL-LHC we are looking at: (i) further reducing the power deposition; (ii) improving the cooling; (iii) using high Curie point ferrites. |
|---|---|---|
| Crab cavities [113] | In the multi kW range if the longitudinal modes overlap with beam harmonic frequencies. | The design should allow detuning the longitudinal modes from multiples of 20 MHz by ~0.5 MHz. |
| Injection protection dump (TDI) [106] | The present design already suffers from beam-induced heating (in the kW range for the injection settings) with nominal LHC parameters due to inefficient cooling. | The present design of the TDI is not compatible with the HL-LHC parameters and a new design is being studied, with installation foreseen for LS2. |
| Synchrotron radiation monitor (BSRT) [114] | The power deposited in the ferrite absorbers (heated at ~ 250°C to 350°C according to simulations and measurements) during 2012 operation could not be efficiently transferred, leading to damage. | A new design is being studied and installed during LS1. The usability of this design for HL-LHC will need to be assessed after LS1. |

*If the longitudinal modes overlap with beam harmonic frequencies.

Table 2-10: Impedance-induced heat loads for the different types of beam screens (including the effects of the longitudinal weld, two counter-rotating beams, and the magneto-resistance) vs. temperature: values are given for the 25 ns beam and for the 50 ns beam (in parentheses).

| Power loss [W/m] | 20 K | 30 K | 40 K | 50 K | 60 K | 70 K |
|---|---|---|---|---|---|---|
| Q1 (49 mm, 6.9 T) | 0.23 (0.28) | 0.24 (0.31) | 0.28 (0.35) | 0.34 (0.43) | 0.40 (0.51) | 0.47 (0.59) |
| Q2–Q3 (59 mm, 8.3 T) | 0.19 (0.24) | 0.20 (0.26) | 0.23 (0.29) | 0.28 (0.35) | 0.33 (0.42) | 0.38 (0.49) |
| D1 (59 mm, 5.6 T) | 0.17 (0.22) | 0.19 (0.24) | 0.22 (0.28) | 0.27 (0.34) | 0.32 (0.41) | 0.38 (0.48) |
| D2 (42 mm, 4.5 T) | 0.25 (0.32) | 0.27 (0.34) | 0.32 (0.40) | 0.39 (0.50) | 0.47 (0.59) | 0.54 (0.69) |
| Q4 (32 mm, 3.7 T) | 0.34 (0.43) | 0.37 (0.47) | 0.44 (0.56) | 0.54 (0.68) | 0.64 (0.81) | 0.74 (0.94) |
| Q5 (22 mm, 4.4 T) | 0.59 (0.74) | 0.63 (0.80) | 0.72 (0.91) | 0.86 (1.09) | 1.01 (1.28) | 1.16 (1.46) |
| Q6 (17.7 mm, 3.5 T) | 0.79 (0.99) | 0.84 (1.06) | 0.96 (1.22) | 1.14 (1.44) | 1.32 (1.68) | 1.51 (1.91) |
| Q7 (17.2 mm, 3.4 T) | 0.82 (1.03) | 0.87 (1.11) | 1.00 (1.26) | 1.18 (1.49) | 1.37 (1.74) | 1.56 (1.98) |

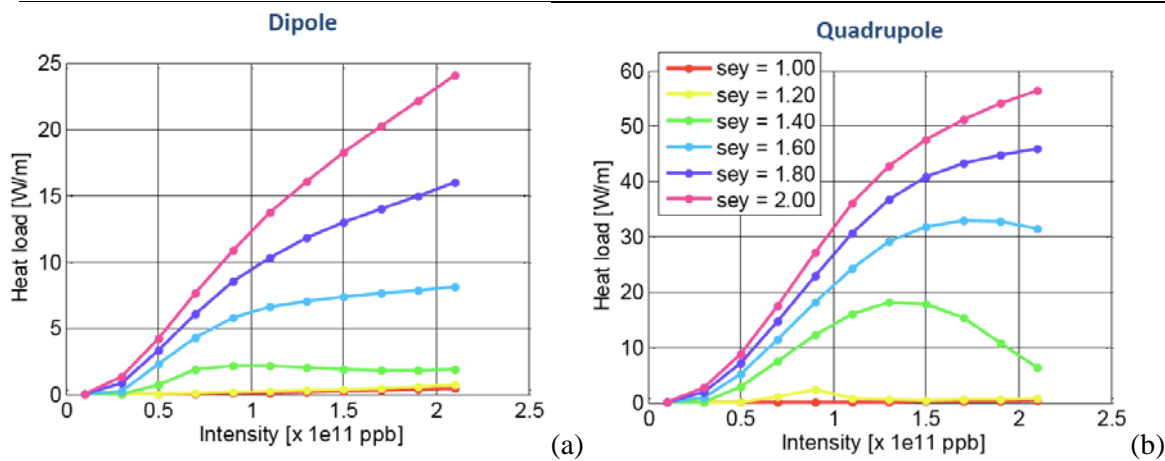

Figure 2-9: Heat load from e-cloud in the arc main magnets as a function of bunch intensity and SEY

Figure 2-10 shows the expected heat load along the triplet in IP1 and IP5 for different SEY values. The least efficient build-up (lower heat load) occurs at the locations of the long-range encounters (vertical dashed

lines). Note that the values in the D1 dipole are comparable to or higher than the values in the quads. The comparison of the heat load from the e-cloud for the current LHC and the future HL-LHC triplets shows that the larger bunch population and larger chamber size lead to a larger heat load by a factor ~3, for the HL-LHC [69] (for the same SEY, a similar energy of multipacting electrons, and a larger number of impacting electrons). For IP2 and IP8, scaling the results also leads to an increase of the heat load from the e-cloud by a factor ~3, but detailed simulations remain to be done. Unlike IP1 and IP5, the cryostats in IP2 and IP8 already include D1 (about 10 m long).

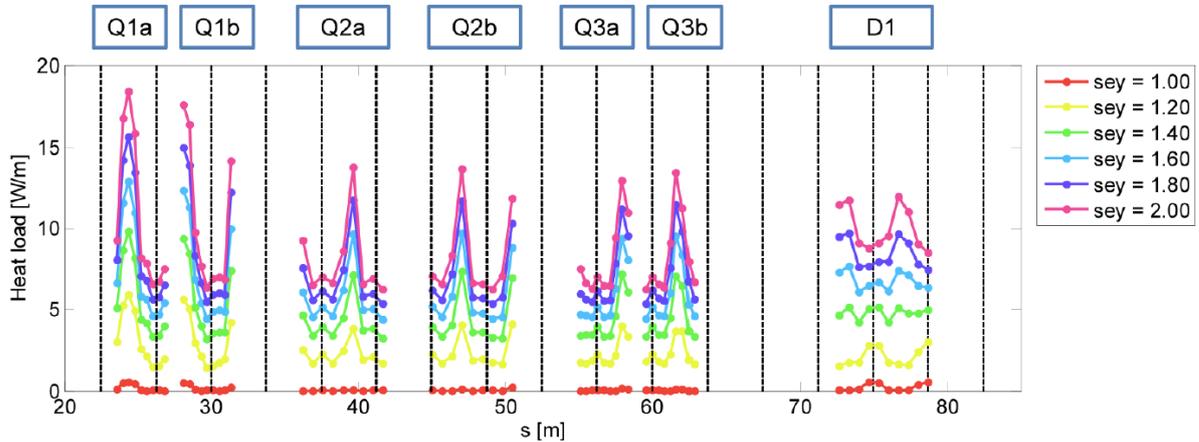

Figure 2-10: Heat load from the e-cloud along the future HL-LHC triplets

Understanding e-cloud build-up in the matching section requires the study of a large number of configurations (beam screen shape and dimensions, magnetic field configuration, beam size, beam position, etc.). Parametric studies will be performed to assess which of these dependencies strongly impact e-cloud build-up. Some preliminary simulations revealed that the beam size and the magnetic field have a small but non-negligible impact. The next step will be to try and disentangle these two effects. The effect of the beam position will then have to be studied in detail.

2.4.4    Luminosity performance

The peak performance at 7 TeV has been estimated in Table 2-1. The estimate of the integrated luminosity requires determining the luminosity evolution during a fill. The beam intensity evolution has been evaluated taking into account burn-off due to luminosity considering a total cross-section of 100 mb [115, 116], and an additional (unknown) source of intensity loss with a lifetime of 200 h (based on experience during 2012 [117]).

The emittance evolution has been determined including: intra-beam scattering (IBS) (based on Run 1 experience, no coupling has been assumed); radiation damping; and an additional (unknown) source of vertical emittance blow-up with a lifetime of 40 h (based on observations during Run 1). A finite difference method in steps of 5 min has been implemented to model the intensity evolution and the evolution of the IBS lifetime as a function of the bunch population. Figure 2-11, Figure 2-12 and Figure 2-13 show the evolution of the main parameters for two cases of levelling, corresponding to a pile-up of 140 and 210, for the standard filling scheme with parameters listed in Table 2-1. In the estimates the worst case scenario (with respect to the head on beam–beam tune spread) of $\beta^*$ levelling in IP1, IP5, and IP8 has been considered. Full compensation of the crossing angle by crab cavities has been included for IP1 and IP5. The crossing angle is assumed to be constant during the fill. Alternative (or complementary) luminosity levelling scenarios include:

- crossing angle variation to increase the geometric reduction factor at the beginning of the fill;
- crab cavity RF voltage variation to have a partial crossing angle compensation at the beginning of the fill;
- dynamic bunch length reduction;
- controlled variation of the transverse separation of the two colliding beams.

Options 1 and 2 have the disadvantage of increasing the line pile-up density at the start of the fill.

The performance estimates for the two cases considered (pile-up 140 and 210) are listed in Table 2-11. The parameters used for the estimates of the HL-LHC integrated performance are listed in Table 2-12. The operation at the higher pile-up limit is appealing because it allows higher integrated luminosity while keeping the optimum fill length to values already obtained in 2012; however, in this case, the maximum pile-up density exceeds the target pile-up density limit of 1.3 events/mm/crossing. This could be reduced by increasing the bunch length. A more elegant solution could be provided by the crab-kissing scheme [16] and/or by the implementation of flat-beam optics with a smaller crossing angle (provided that long-range effects can be kept under control, perhaps by the implementation of long-range compensation schemes). These alternative scenarios will be described below.

Table 2-11: Integrated performance estimate for levelling scenarios at pile-up levels of 140 (PU 140) and 210 (PU 210) events/crossing, respectively.

|  | Levelling time [h] | Optimum fill length [h] | Integrated Lumi [fb$^{-1}$/y] for $\eta$ = 50%, optimum fill length IP1/IP5/IP8/IP2 | Maximum mean pile-up density [events/mm/crossing] in IP1/IP5 | Maximum mean pile-up [events/crossing] in IP1/IP5 |
|---|---|---|---|---|---|
| PU 140 | 8.1 | 9.4 | 261/8.9/0.105 | 1.25/140 | 140 |
| PU 210 | 4.25 | 6 | 331/7.8/0.092 | 1.81/210 | 210 |

Table 2-12: Parameters assumed for HL-LHC performance estimate

| **Scheduled physics time for p–p luminosity production/year ($T_{phys}$) [days]** | **160** |
|---|---|
| Minimum turnaround time [h] | 3 |
| Performance efficiency – goal [%] | 50 |
| Pile-up limit IP1/IP5 [events/crossing] | 140/200 |
| Pile-up limit IP8 [events/crossing] | 4.5 |
| Pile-up density limit – IP1/IP5 [events/mm/crossing] | 1.3 |
| Visible cross-section IP1/IP2/IP5 [mb] | 85 |
| Visible cross-section IP8 [mb] | 75 [118] |

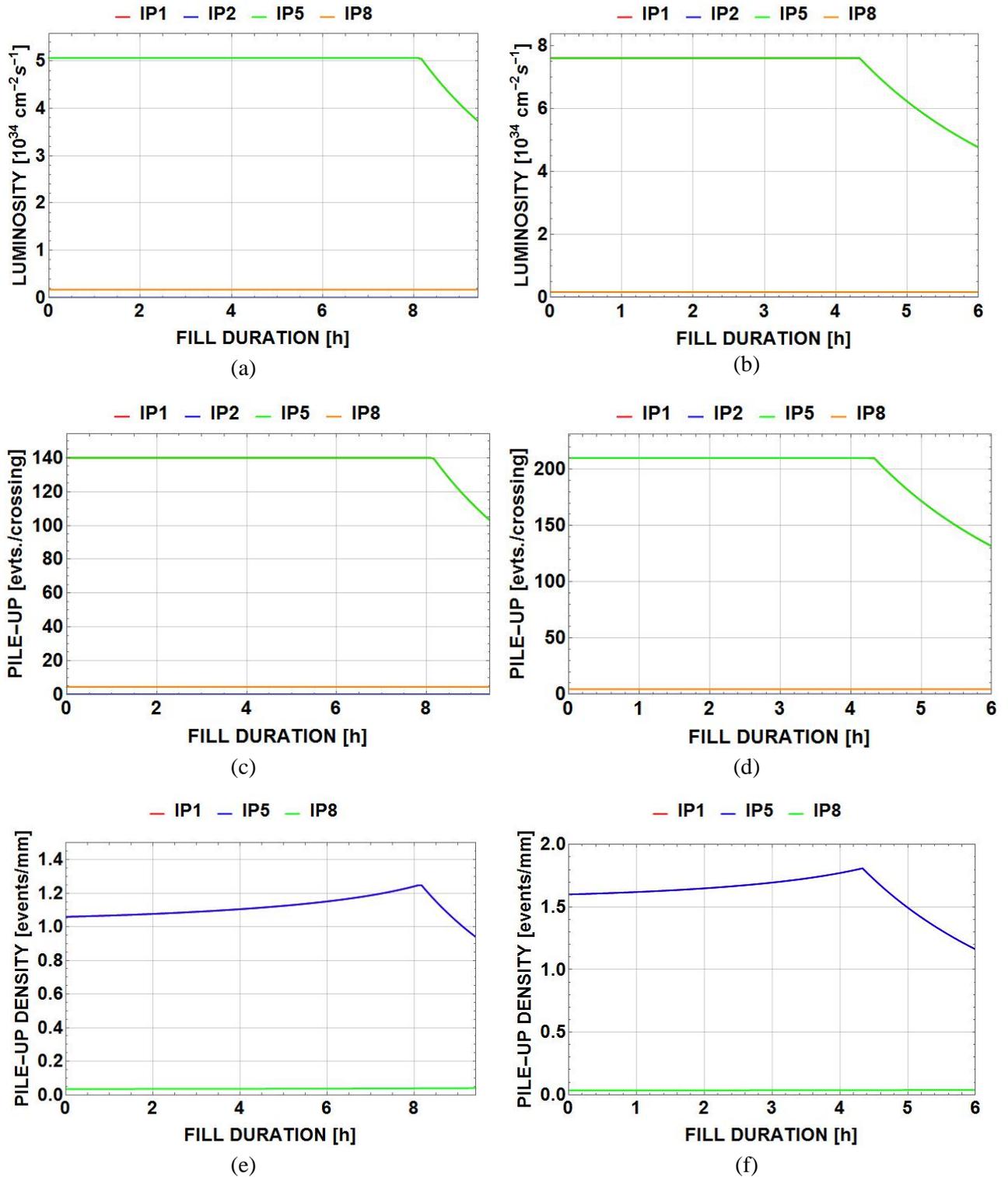

Figure 2-11: Evolution of Luminosity, pile-up and pile-up density assuming levelling at 140 events/crossing (a, c, e) and 210 events/crossing (b, d, f) in IP1 and IP5. The small effects of RF curvature in the crab cavities is not included.

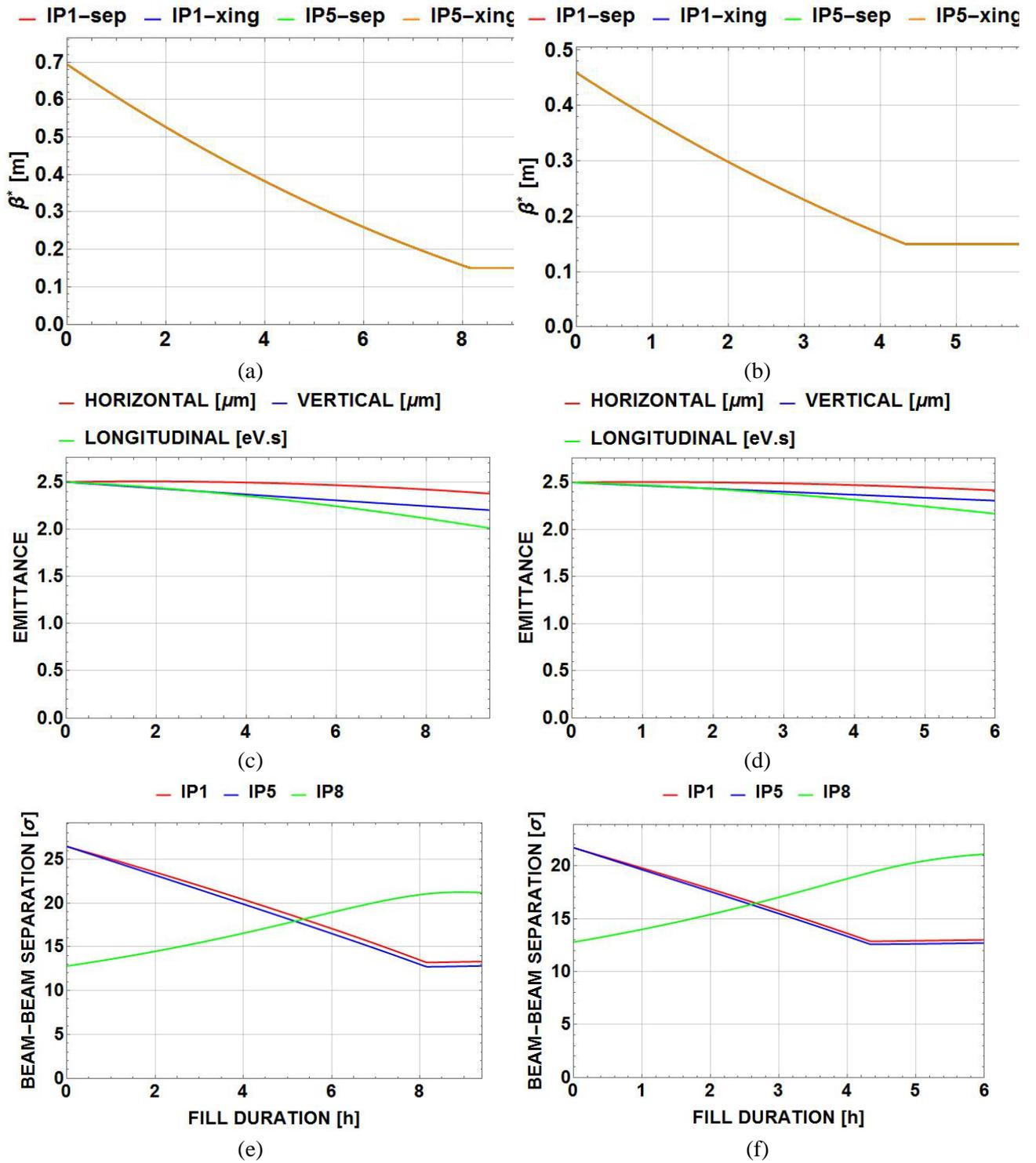

Figure 2-12: Evolution of $\beta^*$, emittance and long-range beam–beam normalized separation ($d_{bb}$) for levelling at 140 (a, c, e) and at 210 events/crossing (b, d, f) in IP1 and IP5.

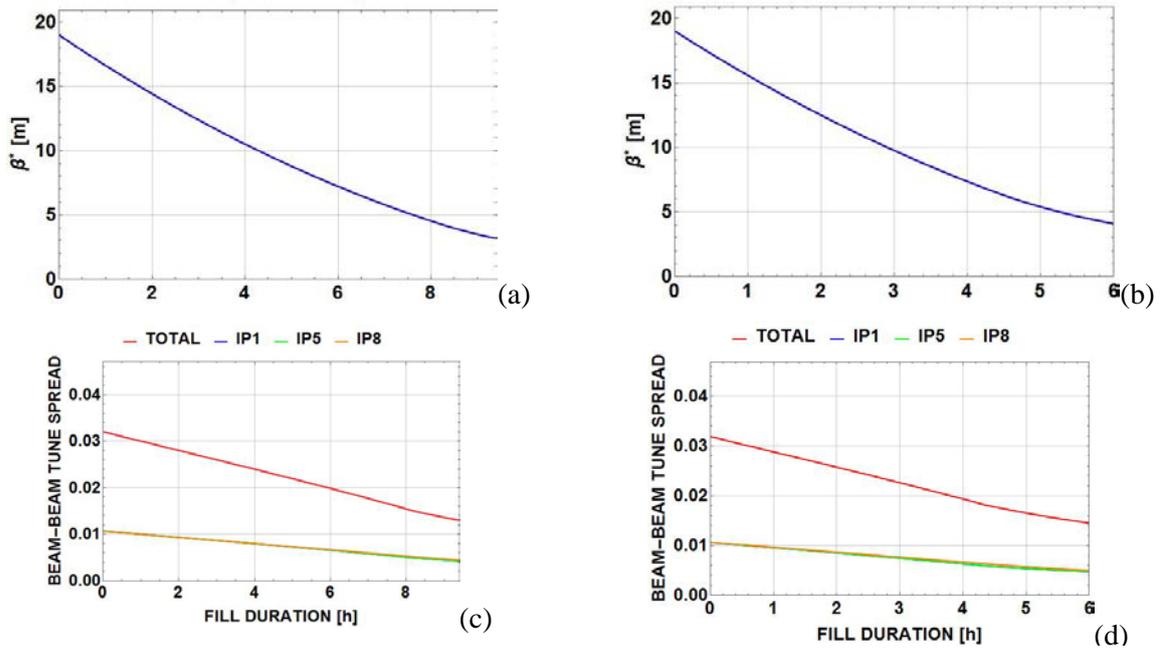

Figure 2-13: Evolution of $\beta^*$ in IP8, head-on beam–beam tune spread assuming levelling at 140 events/crossing (a, c) and at 210 events/crossing (b, d) in IP1 and IP5.

## 2.5 Variants and options

The HL-LHC project includes the study of various alternatives to the present baseline configuration with the aim either of improving the potential performance of the machine or of providing options for addressing possible limitations or changes in parameters (see Figure 2-14).

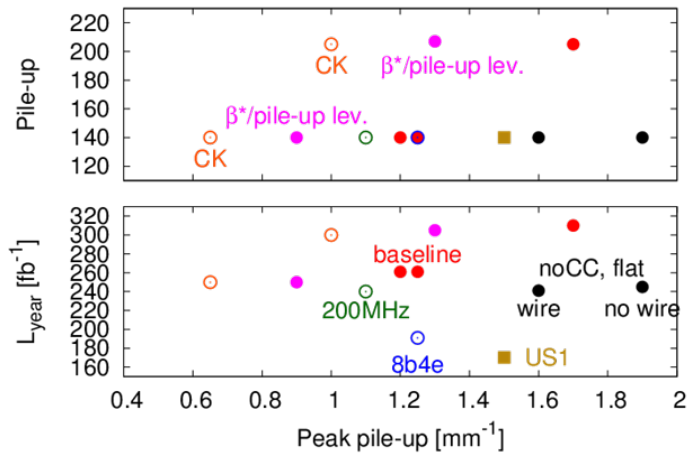

Figure 2-14: Performance expectation of different alternatives. The red markers represent the baseline scenarios (flat or round) for different target of total pile-up (140 and 200 events per crossing). In the case of the crab cavities are absent (black marker) annual integrated luminosity could be recovered by using flat optics for an increase of the peak pile-up density that can be limited with long-range compensators. The crab-kissing scheme (magenta markers) on the other hand offers the target annual luminosity with the means to control the pile-up density and to reduce it considerably with respect to the nominal scheme. Peak pile-up density can also be levelled with $\beta^*$ (magenta markers). If e-cloud effects needs to be mitigated, a 200 MHz RF system or the 8b+4e filling schemes (green and blue markers) could be deployed.

### 2.5.1 Pile-up density management

The crab-kissing scheme ([16, 119] and references therein) would allow either improvement of the data quality by reducing the peak pile-up density at a constant integrated luminosity, or an increase of the integrated luminosity at constant total pile-up or levelled luminosity (see orange markers in Figure 2-14). At present, implementation of the crab-kissing scheme would require changes to the layout of the crab cavities, to allow bunch deflections in different planes. Also, a flat optics with beam–beam long-range compensator would need to be implemented. These changes are not completely compatible with the nominal scheme because the crab cavity voltage is insufficient to provide the required deflections in both planes; however new optics solutions are being explored to keep the crab-kissing scheme and the nominal configuration mutually compatible.

If crab kissing is not available it would still be possible to level peak pile-up density using $\beta^*$ with different total pile-up targets (see magenta markers in Figure 2-14). In this scenario a flat longitudinal profile is assumed to be reachable through use of a second harmonic RF system or phase modulation as in the case of crab kissing. In the absence of flat longitudinal profiles one could expect an increase in the peak pile-up of about 10%.

### 2.5.2 Alternative scenarios

A compact 200 MHz RF system could be installed ([79] and references therein) in the LHC, potentially allowing an increase in the injected intensity from the SPS, although further studies are needed to confirm what could be achieved. Such a system, together with a 400 MHz system, could offer the means to increase the bunch length (e.g. 15 cm) to reduce electron cloud effects, reduce IBS growth rates, and provide flat longitudinal bunch charge density (see green markers in Figure 2-14).

If the e-cloud severely limits the beam current with 25 ns bunch spacing, the 8b+4e [79] filling scheme would allow comparable beam current at the cost of lower levelled luminosity with constant total pile-up. The 8b+4e scheme can provide about $2.3 \times 10^{11}$ p/b with 1900 bunches [3], halfway between the corresponding 50 ns and 25 ns configurations, and therefore resulting in a performance reach between those two extremes (see blue markers in Figure 2-14).

In the absence of crab cavities, it may still be possible to implement some measures to limit the loss of performance [20]. By increasing $\beta^*$ in the crossing plane and decreasing it in the separation plane, one could limit the impact of the geometric reduction factor thanks to the reduction of the necessary crossing angle (see rightmost black marker in Figure 2-14). However, a large crossing angle in units of σ is still needed (12 σ for the lower $\beta^* = 30$ cm) because of the partial loss of the IR1/IR5 long-range beam–beam (LRBB) interaction compensation (i.e. passive compensation from the non-symmetric alternating crossing between IP1 and IP5). However, an LRBB compensator could potentially allow 10 σ separation, therefore restoring the luminosity compared to the nominal scenario at the cost of some increase in pile-up density (see the leftmost black marker in Figure 2-14). Similarly, a staged upgrade scenario for which the replacement of the matching section is postponed (see Refs. [120, 47] and references therein) could still benefit from flat beam optics, although with a limited reach of $\beta^*$ in the non-crossing plane (about 20 cm) but with the potential of reaching about 170 fb$^{-1}$ per year (see dark yellow markers in Figure 2-14). In this context, a design for a movable TAXN that could be adapted to different $\beta^*$ values would offer the best radiation protection for the downstream elements in all these scenarios, regardless of the chosen optics configurations.

## 2.6 The HL-LHC as a nucleus–nucleus collider

The LHC's second major physics programme provides nucleus–nucleus collisions to ALICE, CMS and ATLAS, and proton–nucleus collisions to these three experiments and, in addition, to LHCb. The overall goal of the programme is ultimately to accumulate 10 nb$^{-1}$ of Pb–Pb luminosity during the whole LHC operating period after Run 2 [80, 121]. The p–Pb requirement will be for approximate equivalence in terms of integrated nucleon-pair luminosity [80, 122, 123]. The heavy-ion programme may also require short p–p runs at specific energies to provide reference data; the luminosity requirement will be similar (see 16.6). Nuclei other than

$^{208}$Pb$^{82+}$ have not been requested by the experiments but remain as possible options with potential performance to be evaluated.

The heavy-ion luminosity upgrade aims at increasing integrated rather than peak luminosity and is therefore focused mainly on injecting the maximum beam current possible. With the expected upgrade to remove the event rate limit of the ALICE experiment, luminosity levelling will no longer be a necessity but may be employed to mitigate the rapid luminosity decay due to the large electromagnetic cross-sections [80, 124]. Low values of $\beta^*$ are required at three or four interaction points so the ATS will not be used. The main elements of the heavy-ion luminosity upgrade should be in place a few years before those of the proton–proton part of the project.

Upgrades to the heavy-ion injector chain [125] would normally aim to increase the number of bunches and the intensity per bunch, but these two quantities are not independent. Injecting long trains from the SPS lengthens the injection plateau in the SPS, subjecting some bunches to higher losses from the effects of intra-beam scattering, space charge, and RF noise [126]. On the other hand, injecting a larger number of short trains from the SPS increases the average bunch intensity but leaves more gaps in the LHC bunch train and increases the LHC's injection time, reducing overall efficiency and subjecting some bunches to more emittance growth at LHC injection. In all cases, there is a broad distribution of bunch parameters in collision in the LHC. Optimization of the injection and filling schemes has to take all these interdependencies into account [126] and will likely have to be done anew each year as a function of injector performance.

The present performance estimates are nevertheless based on an injection scheme that assumes that the maximum 12 PS batches of four bunches are assembled into a batch in the SPS, with a 50 ns bunch spacing achieved by slip-stacking. This is repeated 26 times to assemble a train of 1248 bunches in the LHC (practical filling scheme details will reduce this by some percent, depending on the experiment) and yielding a distribution of individual bunch-pair luminosities at the start of colliding beams as shown in Figure 2-15. Simulation of the evolution of these individual bunches, taking into account luminosity burn-off, IBS, and radiation damping [126] leads to the total luminosity shown in Figure 2-16. Depending on the turnaround time (between beam dump and the next declaration of stable beams for physics), the fill length can be optimized to give the ideal average daily luminosity shown.

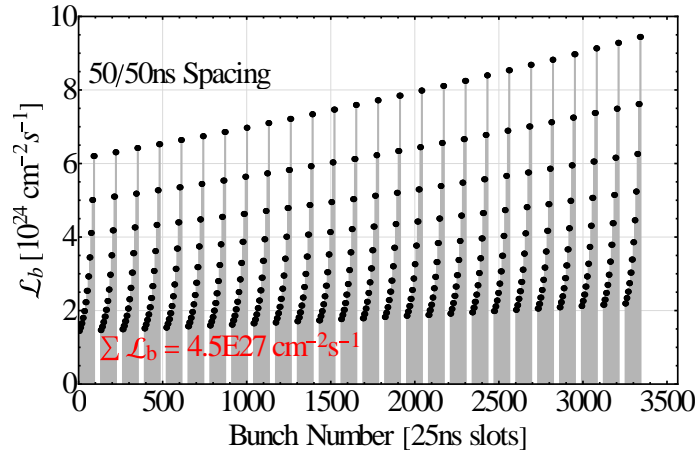

Figure 2-15: Initial luminosity for each colliding bunch pair along the full train in the LHC

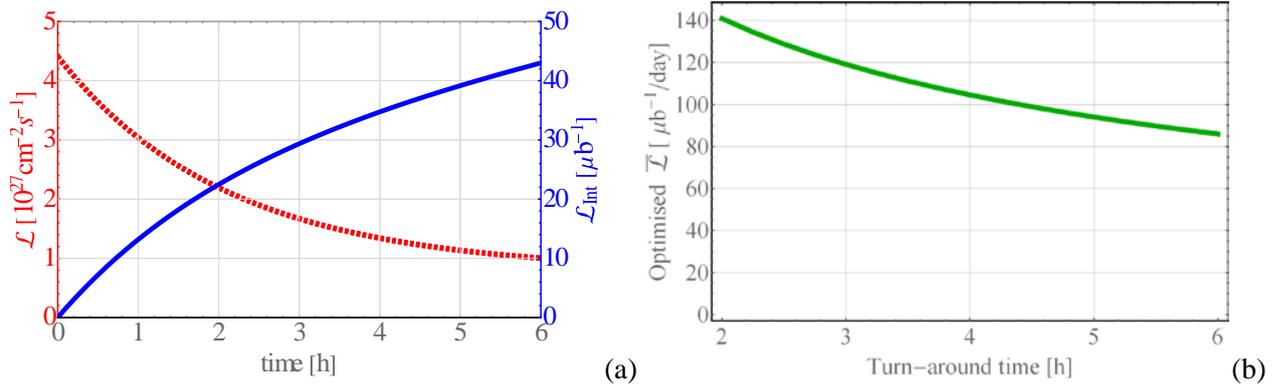

Figure 2-16: (a) Total luminosity (red) and integrated luminosity (blue) during a fill starting with the bunch-pair luminosities shown in Figure 2-15. (b) Average luminosity per day as a function of turnaround time (dump to next stable beams) when fill lengths are optimized, with lengths varying between 3 h and 6 h, with the luminosity dependence shown in the left plot.

The rapid luminosity decay is due to the large cross-sections of electromagnetic processes in the collisions [126, 127]. The peak luminosity is expected to be substantially above the quench limit from losses due to the bound-free pair-production process; and new collimators are foreseen in the dispersion suppressors to absorb these secondary beams emerging from the interaction points (see Chapter 5, Section 5.2). Intensity limitations may also arise from losses due to collimation inefficiency, which is higher for ion beams, due to the more complicated nuclear interactions with collimators [128, 129].

The 50 ns bunch spacing introduces close parasitic beam–beam encounters near to the ALICE experiment, which may require the half-crossing angle to be increased beyond the 60 µrad limit imposed for optimum operation of the zero-degree calorimeters. The minimum acceptable at the low Pb bunch charge will be determined empirically [80]. The crossing angles for ATLAS and CMS are unrestricted and can be taken over from proton operation.

The principal beam parameters determining the luminosity are summarized in Table 2-13. Other parameters will be similar to those given in Ref. [130]. Further potential gains in luminosity may come from improved injector performance and, possibly, a cooling system [131] in the LHC.

Table 2-13: Average values of principal beam parameters at start of physics

| Parameter | Value |
| --- | --- |
| Number of bunches per beam | 1248 |
| Normalized transverse emittance (average) | 1.6 µm |
| Optical function at interaction point | 0.5 m |
| Crossing angle at ALICE experiment | 60 µrad |
| Bunch population (average) | $1.04 \times 10^8$ |
| Bunch length | 0.1 m |
| Peak luminosity | $4.5 \times 10^{27}$ cm$^{-2}$ s$^{-1}$ |

## 2.7 Acknowledgements

The volunteers of the LHC@home project are warmly acknowledged for their generous contribution of CPU time for the numerical simulations of dynamic aperture. Furthermore, we would like to express our gratitude to E. McIntosh for his continuous support for the numerical simulations and the SixTrack code.